%% file: iclr2026_conference.tex
\documentclass{article} 
\usepackage{iclr2026_conference,times}

\input{math_commands.tex}

\usepackage[T1]{fontenc}
\usepackage{hyperref}
\usepackage{url}

\usepackage{booktabs}
\usepackage{nicefrac}
\usepackage{microtype}
\usepackage{graphicx}
\usepackage{pifont}
\usepackage{algorithm}
\usepackage{algpseudocode}
\usepackage{wrapfig}
\usepackage{xspace}
\usepackage[table]{xcolor}
\usepackage[most,minted,listings]{tcolorbox}
\usepackage{makecell}
\usepackage{subcaption}
\usepackage{lipsum}
\usepackage{listings}
\usepackage{cleveref}
\usepackage{colortbl}
\usepackage{soul}
\usepackage{array}
\usepackage{enumitem}
\usepackage{multirow}
\usepackage{dsfont}
\usepackage{caption}

\newcommand{\sys}{HarnessLLM\xspace}

\definecolor{transgray}{gray}{0.9}
\definecolor{rulegray}{rgb}{0.7,0.7,0.8}
\definecolor{lightblue2}{RGB}{100, 181, 205}
\definecolor{lightbluebg}{rgb}{0.96,0.98,1.0}

\newcommand{\e}[1]{{$#1$}}

\title{HarnessLLM: Automatic Testing Harness Generation via Reinforcement Learning}

\author{Yujian Liu$^1$,
Jiabao Ji$^1$,
Yang Zhang$^2$,
Wenbo Guo$^1$,
Tommi Jaakkola$^3$,
Shiyu Chang$^1$ \\
$^{1}$UC Santa Barbara \quad
$^{2}$MIT-IBM Watson AI Lab \quad
$^{3}$MIT CSAIL \\
\texttt{\{yujianliu,jiabaoji,henrygwb,chang87\}@ucsb.edu}, \\
\texttt{yang.zhang2@ibm.com},\quad\texttt{tommi@csail.mit.edu}
}

\iclrfinalcopy 
\begin{document}

\maketitle

\begin{abstract}
Existing LLM-based automatic test generation methods mainly produce input and expected output pairs to categorize the intended behavior of correct programs. Although straightforward, these methods have limited diversity in generated tests and cannot provide enough debugging information.
We propose HarnessLLM, a two-stage training pipeline that enables LLMs to write harness code for testing. Particularly, LLMs generate code that synthesizes inputs and validates the observed outputs, allowing complex test cases and flexible output validation such as invariant checking. 
To achieve this, we train LLMs with SFT followed by RLVR with a customized reward design. 
Experiments show that HarnessLLM outperforms input-output-based testing in bug finding and testing strategy diversity. 
HarnessLLM further benefits the code generation performance through test-time scaling with our generated test cases as inference-phase validation. Our code is available at \url{https://github.com/UCSB-NLP-Chang/HarnessLLM.git}.
\end{abstract}

\input{section/1_intro}
\input{section/2_related_work}
\input{section/3_method}
\input{section/4_experiment}
\input{section/5_conclusion}
\clearpage
\input{section/6_addition}

\bibliography{iclr2026_conference}
\bibliographystyle{iclr2026_conference}

\clearpage

\appendix
\input{section/appendix}

\end{document}

%% file: math_commands.tex
\usepackage{amsmath,amsfonts,bm}

\def\eqref#1{equation~\ref{#1}}

\def\1{\bm{1}}

\DeclareMathAlphabet{\mathsfit}{\encodingdefault}{\sfdefault}{m}{sl}
\SetMathAlphabet{\mathsfit}{bold}{\encodingdefault}{\sfdefault}{bx}{n}



%% file: section/1_intro.tex
\section{Introduction}
\label{sec:intro}

Large language models (LLMs) have demonstrated remarkable proficiency in code-related tasks, including code generation, completion, and even resolving software engineering issues through tool use~\cite{chen2021evaluatinglargelanguagemodels,Li_2022_alpha,openai2024openaio1card,deepseekai2025deepseekr1incentivizingreasoningcapability,jimenez2024swebench,lipatchpilot}. 
However, compared to these code generation tasks, automatic testing and debugging AI-generated programs have received comparatively little attention, even though comprehensive test suites are critical for ensuring the correctness and robustness of the AI-generated code~\cite{chen2024teaching,prasad2025learninggenerateunittests,sinha2025languagemodelsfalsifyevaluating,he2025hardtestssynthesizinghighqualitytest,zhang-etal-2023-self}.

\begin{figure*}[t]
\centering
    \begin{minipage}[t]{0.45\textwidth}
        \vspace{0pt}\centering
        \includegraphics[width=\linewidth]{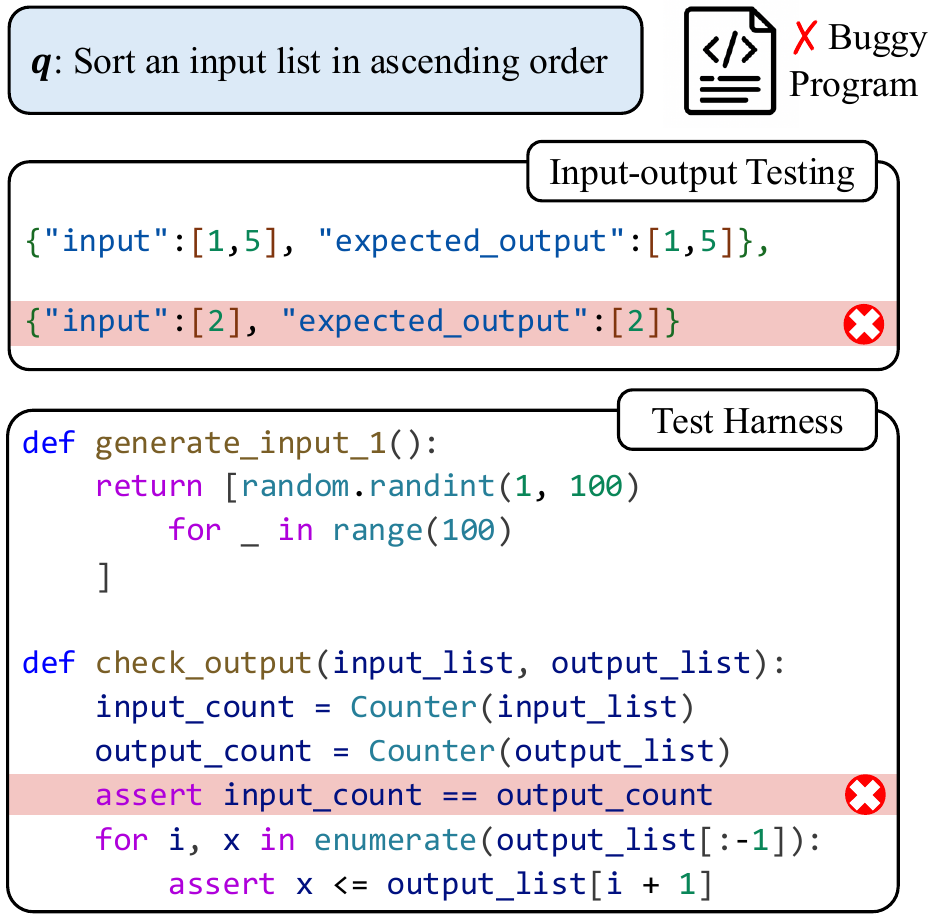}
        \caption{Comparison between input-output pairs (top) and test harness (bottom).}
        \label{fig:intro_example}
    \end{minipage}
    \hspace{0.01\textwidth}
    \begin{minipage}[t]{0.48\textwidth}
        \vspace{0pt}\centering
        \includegraphics[width=\linewidth]{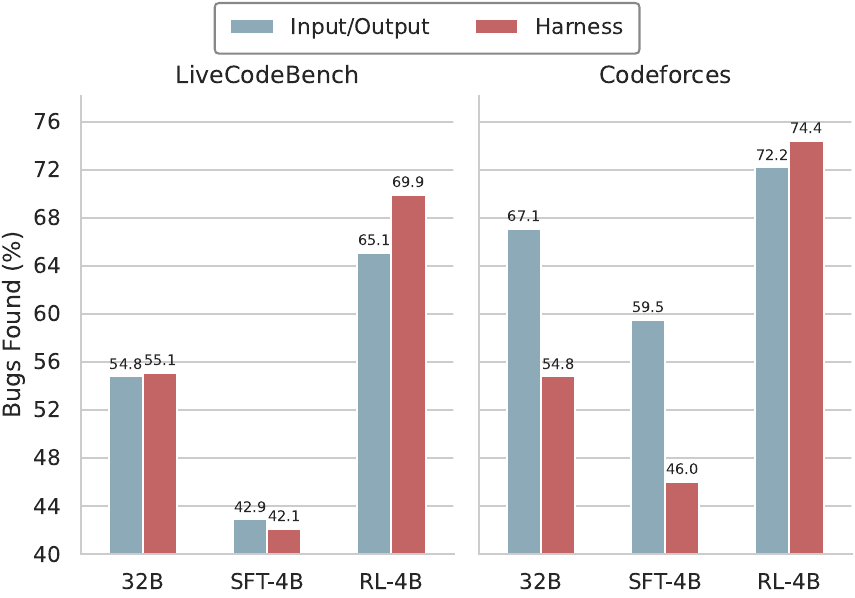}
        \caption{Percentage of found bugs (average of 8 runs, higher is better) for two strategies with different models.}
        \label{fig:intro_perf}
    \end{minipage}
\vspace{-5mm}
\end{figure*}

Existing works in automatic testing mainly prompt the language model to generate input–output pairs that characterize the intended behavior of correct programs~\cite{chen2022codetcodegenerationgenerated,prasad2025learninggenerateunittests,zeng2025acecoderacingcoderrl,lin2025learningsolveverifyselfplay}.
As depicted in Figure~\ref{fig:intro_example}, the model produces examples of test inputs alongside their expected test outputs. The target program is then executed on a test input, and the output that the program generates is compared against the corresponding expected test output. A bug is exposed if the two outputs diverge.

Although straightforward, such an input-output test case generation paradigm has two potential drawbacks. \textit{First}, the test inputs generated by the language model tend to be simple and homogeneous, so they may not have sufficient coverage of the sophisticated corner cases that could expose bugs. \textit{Second}, such a paradigm requires that the model generates the correct output by itself, which becomes extremely challenging for complicated programming tasks or for complicated test inputs. In short, the fundamental paradox is that the `tester', \emph{i.e.,} the language model that generates test cases, is often much weaker than the `testee', \emph{i.e.,} the program, in accomplishing the complex programming tasks (otherwise, the language models would not have to rely on code generation to solve these tasks).

In this paper, we explore a novel debugging paradigm that could resolve this paradox -- LLM-based \emph{test harness generation}. Rather than letting the language model directly generate input-output pairs, we prompt it to write executable code, a matching rival to the testee, to generate test inputs and validate target program outputs. In this way, both aforementioned drawbacks can be addressed simultaneously. On the input side, executable code can easily generate various richly structured, diverse, and complicated inputs. On the output side, executable code opens up many possibilities to validate target program outputs. It can \ding{182} directly generate hardcoded expected output, as does the input-output paradigm, or \ding{183} write a reference program to compute the expected output, or, most interestingly \ding{184} assert output properties and requirements.
As shown in Figure~\ref{fig:intro_example}, for a program that sorts a list of input integers, the LLM first writes an input generator, \texttt{generate\_input\_1}, to generate random lists, which are fed to the target program for execution. The returned outputs are then validated by an LLM-defined function, \texttt{check\_output}, which checks that the result is sorted and preserves the original integers.
With programmatic input generation and output validation, testing harnesses can support complex invariant checking and stress testing, enabling more comprehensive testing and detection of deep logical bugs.

However, the key challenge of the test harness generation paradigm is that even the strong coding LLMs are not inherently capable of test code generation, which requires different skills than generating code for programming tasks: The former requires understanding the given program’s logic, control and data flow, designing proper stress tests, and writing validation logic, while the latter is mainly about writing code to fulfill the required functionality. 
To demonstrate this, our initial experiment compares the bug‐finding rates of the input–output strategy versus test harness generation on the \textsc{LiveCodeBench} and \textsc{Codeforces} datasets \cite{jain2025livecodebench, penedo2025codeforces}, using a strong reasoning model \texttt{Qwen3-32B} \cite{yang2025qwen3technicalreport}. 
Surprisingly, direct prompting for test harnesses does not yield better bug finding capabilities (Figure~\ref{fig:intro_perf}).

To close this gap, we propose \sys, a two‐stage training pipeline combining supervised fine‐tuning (SFT) with reinforcement learning (RL) with customized reward functions. 
First, we collect SFT data by prompting \texttt{Qwen3-32B} and filtering for harnesses that successfully expose a bug. 
We warm up a smaller model (\emph{e.g.,} \texttt{Qwen3-4B}) with SFT on collected data. 
The purpose of this stage is to provide a reasonable starting point for reinforcement learning, which improves RL's training efficiency. 
Second, we further train the SFT model using RL with our customized verifiable outcome reward.
Here, we assume access to a ground-truth program during training. To encourage the model to generate valid harnesses, we first give a zero reward to generated tests that trigger compilation or runtime errors on the ground-truth program. Then, we design rewards to incentivize the model to generate effective tests that crash the target programs. Specifically, a positive reward is assigned when the ground-truth program can pass the generated tests but the target program fails, indicating that the test harness correctly identifies bugs in the target program.
We train the model to maximize the expected reward using the GRPO algorithm \cite{shao2024deepseekmathpushinglimitsmathematical}. 
The RL training can further strengthen the model's capabilities to generate effective test harnesses and improve the model's generalizability.

We train on two base models (\texttt{Qwen3-4B} and \texttt{Llama3.2-3B} \cite{grattafiori2024llama3herdmodels}) and evaluate on three benchmarks containing buggy programs. Experiments show that our model outperforms all baselines, including the off-the-shelf \texttt{Qwen3-32B} and another model that is also trained with RL but only generates input-output pairs (Figure \ref{fig:intro_perf} presents an overview).
Moreover, the learned harness generator generalizes to code produced by unseen models and can be used for improving code generation performance. 
Specifically, using the execution results of generated test cases to select the best out of 8 responses improves \texttt{Qwen3-32B}'s performance from 63.5\% to 69.5\% on \textsc{LiveCodeBench} \cite{jain2025livecodebench}.
To the best of our knowledge, \sys is~\emph{the first LLM-based testing harness generation that enables comprehensive testing and benefits competitive programming tasks}.

We summarize our contributions as follows:
\begin{itemize}[noitemsep, topsep=0pt]
  \item We propose harness-based automatic program testing, a new debugging paradigm with richer context and more diverse testing cases beyond input-output checks.
  \item We design a pipeline with SFT and RL to train LLMs to write effective test harnesses.
  \item We trained specialized reasoning models using \sys, comparing their effectiveness with SOTA LLMs, and demonstrating their utility in code generation.  
\end{itemize}

%% file: section/2_related_work.tex
\section{Related Works}

\paragraph{Automatic Test Case Synthesis.}
Test cases are crucial in evaluating code correctness. While many established benchmarks rely on manually written test cases~\cite{chen2021evaluatinglargelanguagemodels, austin2021programsynthesislargelanguage, hendrycks2021measuring}, this process is labor-intensive and does not scale well. To address this limitation, a variety of automatic test case synthesis methods have been proposed. Traditional approaches leverage programming language techniques to explore the input space and cover diverse execution paths~\cite{puspitasari2023analysis, forgacs2024modern, guo2024optimal, reid1997empirical}. Although these techniques improve input coverage, they often fall short in capturing code semantic relationships and complex control flows, which can lead to undetected failures during runtime.
Recently, LLMs have been used to synthesize test cases by prompting them to generate both inputs and expected outputs~\cite{yuan2024manualtestsevaluatingimproving,chen2024chatunitestframeworkllmbasedtest,han2024archcodeincorporatingsoftwarerequirements,li2024largelanguagemodelstest,guzu2025large,xiong2023program,wang2025co,cao2025llmsgeneratereliabletest,wang2025codecontestshighqualitytestcase}.
Despite their strong code understanding capabilities, LLMs still struggle to consistently generate correct outputs, especially when the code is complex. 
In this work, we propose a novel paradigm that shifts from output prediction to execution-based validation. Our \sys programmatically generates inputs and validates outputs, expanding the design space of test cases.

\paragraph{Reinforcement Learning with Verifiable Rewards.}
Reinforcement learning has shown great potential in improving LLM abilities in many domains requiring heavy reasoning, such as math problem solving~\cite{deepseekai2025deepseekr1incentivizingreasoningcapability,kimiteam2025kimik15scalingreinforcement,shao2024deepseekmathpushinglimitsmathematical,yu2025dapoopensourcellmreinforcement,hou2025thinkprune0}, code generation~\cite{le2022coderl0, el2025competitive, code-r1}, and robotic control~\cite{chu2023accelerating, ji2025collision0}.
In this work, we use RL to improve LLMs' test case generation abilities. By designing a customized reward that judges whether the generated test cases can differentiate between correct and buggy programs, we train LLMs to learn the reasoning skills required to write effective test cases.

%% file: section/3_method.tex
\section{Methodology}

\subsection{Problem Formulation}
Formally, let \e{\bm q} be the description of a programming problem with input space \e{\mathcal{I}} and output space \e{\mathcal{O}}. Denote \e{f, g: \mathcal{I} \to \mathcal{O}} as two programs for this problem, where \e{f} is a potentially buggy implementation that is under testing, and \e{g} is a ground-truth implementation for the problem. We say \e{f} has logical bugs if for some \e{\bm x \in \mathcal{I}, f(\bm x) \neq g(\bm x)}. In other words, \e{\bm x} triggers the divergent behaviors of the buggy and reference programs. Therefore, an automatic debugging method generally contains two steps: generating inputs that can potentially trigger the bug and comparing the target program's output with the reference output.

However, in most real-world situations, the ground-truth implementation \e{g} is not available, which necessitates an approximate verifier to validate the output of \e{f}. Denote this verifier as \e{v: \mathcal{I} \times \mathcal{O} \to \{0, 1\}}, where \e{v(\bm x, \bm y)=1} indicates that output \e{\bm y} on input \e{\bm x} is deemed correct. Our goal in this paper is to train an LLM for automatic debugging that, given \e{\bm q} and \e{f}, emits both a set of inputs \e{\{\bm x_i\}_{i=1}^N} and a corresponding verifier \e{v}.
Note that we mainly focus on finding \textit{logical bugs} in a target program, \emph{i.e.,} deviations from intended behavior, and leave security vulnerability for future work. 

\paragraph{Challenge of Input-Output Testing.}
The input-output testing can be considered as having a simple verifier that compares the program's output with the expected output. Specifically, the model generates a set of pairs \e{\{(\bm x_i, \hat{\bm y_i})\}_{i=1}^N}, where \e{\hat{\bm y_i}} is the expected output for input \e{\bm x_i}. The verifier is then an indicator function \e{v(\bm x_i, f(\bm x_i)) = \mathds{1}(f(\bm x_i) = \hat{\bm y_i})}. However, this simple verifier requires the model itself to come up with a correct expected output, which limits the complexity of test cases. In the following, we propose a framework that generates test harnesses to address this challenge.

\subsection{Generating Test Harness for Debugging}
\label{subsec:harness}
We propose instead that the LLM writes a test harness code that synthesizes inputs and programmatically checks outputs. 
Having harnesses can help produce more diverse testing cases and provide more valuable feedback when the program crashes.  
Specifically, our framework consists of three steps.

\noindent
\textbf{\textit{Step 1: Generate Input.}}
The model implements a set of input generators, each returning a list of inputs for the program (\emph{e.g.,} \texttt{generate\_input\_1()}). By leveraging loops or random functions, the LLM can craft rich test inputs, which would be difficult to get with hardcoding.

\noindent
\textbf{\textit{Step 2: Execute.}}
Each generated input is fed to the program \e{f}, and the resulting output is captured.

\noindent
\textbf{\textit{Step 3: Validate Output.}}
A model-implemented function \texttt{check\_output(input,output)} is used to validate the correctness of each captured output. The model can use various ways for validation, such as checking specific invariants or comparing with output from a brute-force implementation. This output checker uses assertions to check correctness, and a bug is reported if the assertions fail for \textit{any} pair of generated input and captured output.

Figure \ref{fig:sample} shows a complete example of model generation for this process, and Figure \ref{fig:harness-io} shows the detailed prompt we use.

\begin{figure}
  \centering
  \includegraphics[width=0.9\linewidth]{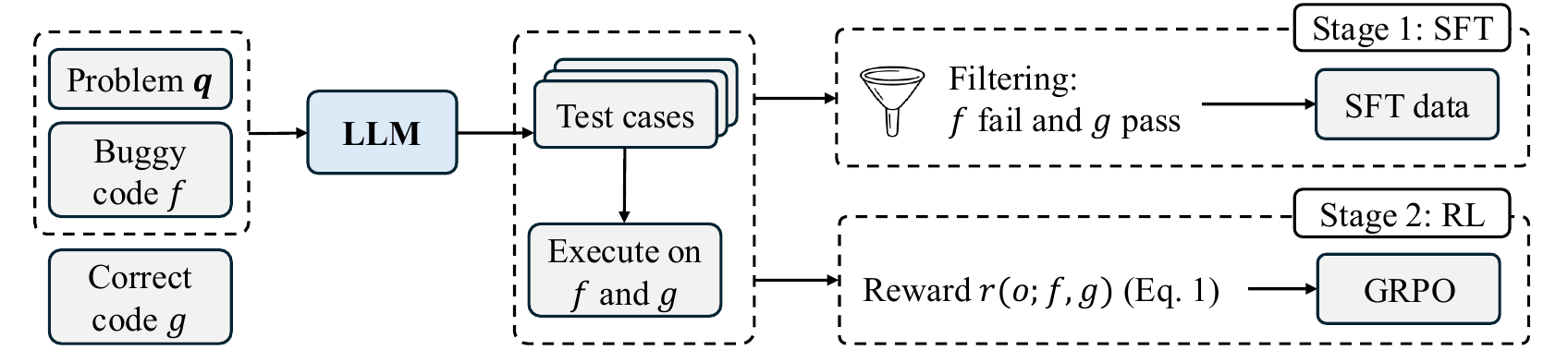}
    \caption{Overview of our training pipeline.}
    \label{fig:training}
    \vspace{-2mm}
\end{figure}

\subsection{Improving Test Harness via RLVR}
\label{subsec:rl}
Despite the promise, we found off‐the‐shelf LLMs struggle to generate effective harnesses. To remedy this, we design a two‐stage training pipeline to improve their performance. Figure \ref{fig:training} depicts an overview of our pipeline.

\paragraph{Stage 1: SFT Warm-Up.}
We prompt \texttt{Qwen3-32B} to generate test harnesses as described in Section \ref{subsec:harness}. The model response contains a long reasoning chain and a final code block. We execute the harnesses against both the target program \e{f} and the ground‐truth program \e{g} and retain only responses for which \e{g} passes but \e{f} fails. We then fine‐tune a smaller model (\emph{e.g.,} \texttt{Qwen3-4B}) with SFT on the filtered dataset. The SFT model has a basic understanding and skills for test harness generation.
Using it as an initialization for RL can improve the learning efficiency of RL, as the early training stage can receive some meaningful positive rewards. 

\paragraph{Stage 2: RL with Verifiable Outcome Reward.}
To further improve the generalizability of the warmed‐up model, we follow recent works to train the model with RL against a verifiable outcome reward \cite{deepseekai2025deepseekr1incentivizingreasoningcapability,lambert2025tulu3pushingfrontiers}. Specifically, for each rollout \e{o} the model generates, let \e{\{\bm x_i\}_{i=1}^N} be the corresponding inputs, we define the following reward function based on the execution results on \e{f} and \e{g}:
\begin{equation}
r(o; f, g) =
\begin{cases}
1, & \text{if }g\text{ passes and }f \text{ fails}; \\
0.1, & \text{if } g \text{ fails\footnotemark or } f \text{ passes, and } \exists\,\bm x_i:\; f(\bm x_i)\neq g(\bm x_i); \\
0, & \text{otherwise}
\end{cases}
\label{eq:reward}
\end{equation}
\footnotetext{Assertion errors in output verifier. All inputs still need to be valid, \emph{i.e.,} do not trigger runtime errors on \e{g}.}
In other words, a reward of 1 is given only when the ground-truth program can pass the test, but not the buggy program, indicating a correct test case. Otherwise, if all inputs are valid (\emph{i.e.,} they do not trigger runtime errors on \e{g}) and at least one input can trigger different outputs for \e{f} and \e{g}, we assign a partial reward of 0.1, which encourages the model to generate bug-exposing inputs. Note that in this case, the input generators work well, but the output verifier generates ineffective assertions, which either fail the correct code \e{g} or do not crash the buggy code \e{f}. Nevertheless, we still assign a partial reward to incentivize the model to generate good inputs. Finally, a reward of 0 is given when no input can expose the bug. Importantly, the requirement that \e{g} has to pass the generated test cases prevents the model from hacking rewards by generating arbitrary invalid tests. We maximize the expected reward using GRPO.

\subsection{Data Collection}
\label{subsec:data}
Both training stages in Section \ref{subsec:rl} require data in the format of a problem description \e{\bm q}, a buggy implementation \e{f}, and a ground-truth implementation \e{g}. To collect such data, we follow prior works \cite{deepcoder2025} to source from existing datasets of coding problems, including TACO \cite{li2023tacotopicsalgorithmiccode}, SYNTHETIC-1 \cite{primeintellect_synthetic1_2025}, LeetCode \cite{xia2025leetcodedatasettemporaldatasetrobust}, and Codeforces \cite{matrixstudio_codeforces_python_2025}. The original solution in the datasets is used as ground-truth program \e{g}, after an additional round of filtering to make sure \e{g} passes all provided test cases of the problem.

To collect the buggy programs \e{f}, we prompt a series of LLMs to solve the problem, including \texttt{Qwen2.5-Coder} 1.5-7B \cite{hui2024qwen25codertechnicalreport} and \texttt{DeepSeek-R1-Distill-Qwen-1.5B} \cite{deepseekai2025deepseekr1incentivizingreasoningcapability}. We only keep programs that satisfy both of the following conditions: \ding{172} The program passes the demo test cases in the problem description; and \ding{173} The program fails on at least one test case of the problem. This makes sure the retained programs are partially correct but still have bugs. We retain at most two buggy programs per problem and select the two that pass the most test cases if multiple programs satisfy the two conditions.

After decontamination against all evaluation data in Section \ref{subsec:setting}, the resulting training set contains 12,043 unique \e{(\bm q, f, g)} triplets. We use all samples for RL training and a subset of 6,805 samples to generate SFT data. Appendix~\ref{append:train-data} details the procedure for our data collection process.

%% file: section/4_experiment.tex
\section{Experiments}
We conduct experiments to validate the effectiveness of HarnessLLM. 
Specifically, we aim to answer two questions: \ding{172} Does our two-stage training pipeline enhance models' ability to write test harnesses? \ding{173} Does harness-based testing outperform input-output testing in identifying bugs?

\subsection{Experiment Setting}
\label{subsec:setting}

\noindent
\textbf{Evaluation Benchmarks.}
We evaluate on three widely used code generation datasets: \textsc{MBPP+} \cite{austin2021programsynthesislargelanguage,liu2023is}, \textsc{LiveCodeBench} \cite{jain2025livecodebench}, and \textsc{Codeforces} \cite{penedo2025codeforces}. We repurpose these datasets for the bug detection task by collecting triplets of problem description, buggy program, and ground-truth program. 
For \textsc{MBPP+}, we directly use the split \textsc{MBPP+Fix (Hard)} in \texttt{UTGen-32B}~\cite{prasad2025learninggenerateunittests}. For \textsc{LiveCodeBench} and \textsc{Codeforces}, we follow the procedure described in Section \ref{subsec:data}. Particularly, we create two dataset variants: \ding{182} \textsc{Seen} version contains buggy programs generated by \texttt{DeepSeek-R1-Distill-Qwen-1.5B}, which is also used to generate our training data. \ding{183} \textsc{Unseen} version contains buggy programs generated by \texttt{Qwen3-14B}, which is never seen during training, and evaluates the generalizability of our models to different code generators. Please see Appendix \ref{append:eval-data} for details of evaluation data.

\noindent
\textbf{Metrics.}
We extend the three standard metrics proposed in \citet{prasad2025learninggenerateunittests} for test harnesses. 
Specifically, \ding{182} \textbf{Good input (GI)} calculates the percentage of responses that have at least one bug-exposing input, \emph{i.e.,} \e{\exists\, \bm x_i: f(\bm x_i) \ne g(\bm x_i)}. This metric purely measures the ability of the input generator. \ding{183} \textbf{Invalid test rate (ITR)} measures the percentage of responses where the ground-truth program fails, \emph{e.g.,} tests that have invalid inputs or incorrect assertions. \ding{184} \textbf{True bug rate (TBR)} measures the percentage of responses that correctly expose the bug, \emph{i.e.,} the ground-truth program passes the tests but the buggy program fails. This metric assesses the \textit{overall performance}.

For each input pair of problem and buggy program, we sample 8 responses and report the average performance of these 8 runs. We follow the official setting of Qwen3 to set the temperature at 0.6 and add a presence penalty of 1.5 \cite{yang2025qwen3technicalreport}. The maximum generation length is set at 32,000.

\noindent
\textbf{Number of Test Cases.}
We allow each model response to contain one or more test cases. Concretely, for input-output testing, each model response could contain multiple pairs of input and expected output. For test harnesses, each response could contain multiple input generators, and each generator could further generate multiple test inputs. In our preliminary experiments, we observe that the number of test cases in each response significantly affects the performance (details in Appendix \ref{append:num-test-detail}). Thus, for the teacher model and SFT models, we report the performance of the best number of test cases. Namely, 1 test case per response for input-output testing and 5 test cases for test harnesses.\footnote{If a response contains more test cases, we only evaluate the first 1 or 5 test cases.} However, restricting the same number of test cases for all problems may be suboptimal. Therefore, during RL training, we allow the model to generate any number of test cases from 1 to 20, and the model learns the optimal number of test cases for each problem through training. The following section reports the performance of the above setting. In Appendix \ref{append:control_test}, we further show results when controlling the number of test cases.

\noindent
\textbf{Baselines.}
We mainly compare with the baseline that generates input-output pairs for testing. For fair comparison, we conduct the same two-stage training as our method. Particularly, we use the same teacher model to generate an equal amount of SFT data, and we use the same reward in Eq.~\ref{eq:reward} for RL training. We additionally report the performance of directly prompting \texttt{Qwen3-32B} with both testing strategies. Finally, we compare with \texttt{UTGen-32B}~\cite{prasad2025learninggenerateunittests}, which also generates input-output pairs but is trained with only SFT without RL.

\noindent
\textbf{Implementation Details.}
We demonstrate the effectiveness of our framework on \texttt{Qwen3-4B} and \texttt{Llama3.2-3B}. For SFT, we train all models for 15 epochs and select the best checkpoint based on the validation performance. For RL, we leverage the Verl training framework \cite{sheng2024hybridflow} and train all models for 500 steps with a batch size of 128. Please see Appendix \ref{append:hyperparam} for detailed training hyperparameters. We parallelize the reward calculation for each rollout across all CPU cores, and on average, it takes 0.06 seconds to execute the test harnesses for each rollout during training. Appendix~\ref{append:dynamics} shows the detailed dynamics in RL training.

\input{tables/main_result}

\subsection{Main Results}
\label{subsec:main-result}

\paragraph{Ability to Find Bugs.}
Table \ref{tab:main-result} shows the performance of \texttt{Qwen3-4B} on finding bugs generated by models that have been seen during training. There are two observations from the table. \textbf{First}, our RL-trained model for test harness generation consistently outperforms the counterpart that generates input-output pairs. Specifically, it achieves better performance on all metrics across all benchmarks, demonstrating the benefits of test harness generation for both input generation and output verification. \textbf{Second}, both RL-trained small models surpass the 32B teacher models, which illustrates the effectiveness of our proposed two-stage training. Interestingly, although test harnesses initially underperform input-output generation on the teacher model and SFT models, our RL training unlocks their advantage and leads to better final performance. Appendix~\ref{append:llama-res} shows the results on \texttt{Llama3.2-3B}, which suggest that our method has better generalizability than input-output testing.

\input{tables/generalization}
\paragraph{Generalizability to Unseen Models.}
We next evaluate our models' ability to debug for models that have never been seen during training. Specifically, we collect buggy programs generated by \texttt{Qwen3-14B}. These buggy programs are different from those in Table \ref{tab:main-result} in two ways: \ding{172} They are from an unseen model and thus may have different distributions for the bugs in the code. \ding{173} They are from a stronger model and pass more test cases, so they contain deeper logical bugs.
Performance shown in Table \ref{tab:generalization} illustrates similar observations as Table \ref{tab:main-result}. Particularly, our RL-trained test harness generators substantially outperform the model that generates input-output pairs. Moreover, our method achieves larger improvements than Table \ref{tab:main-result}. For instance, the relative improvement on \textsc{Codeforces} increases from 3.0\% to 17.0\%. The results show that our models can better generalize to unseen models. It also validates that the improvements of our method are not overfitting to a particular distribution of bugs.

\begin{figure}
  \centering
  \includegraphics[width=0.95\linewidth]{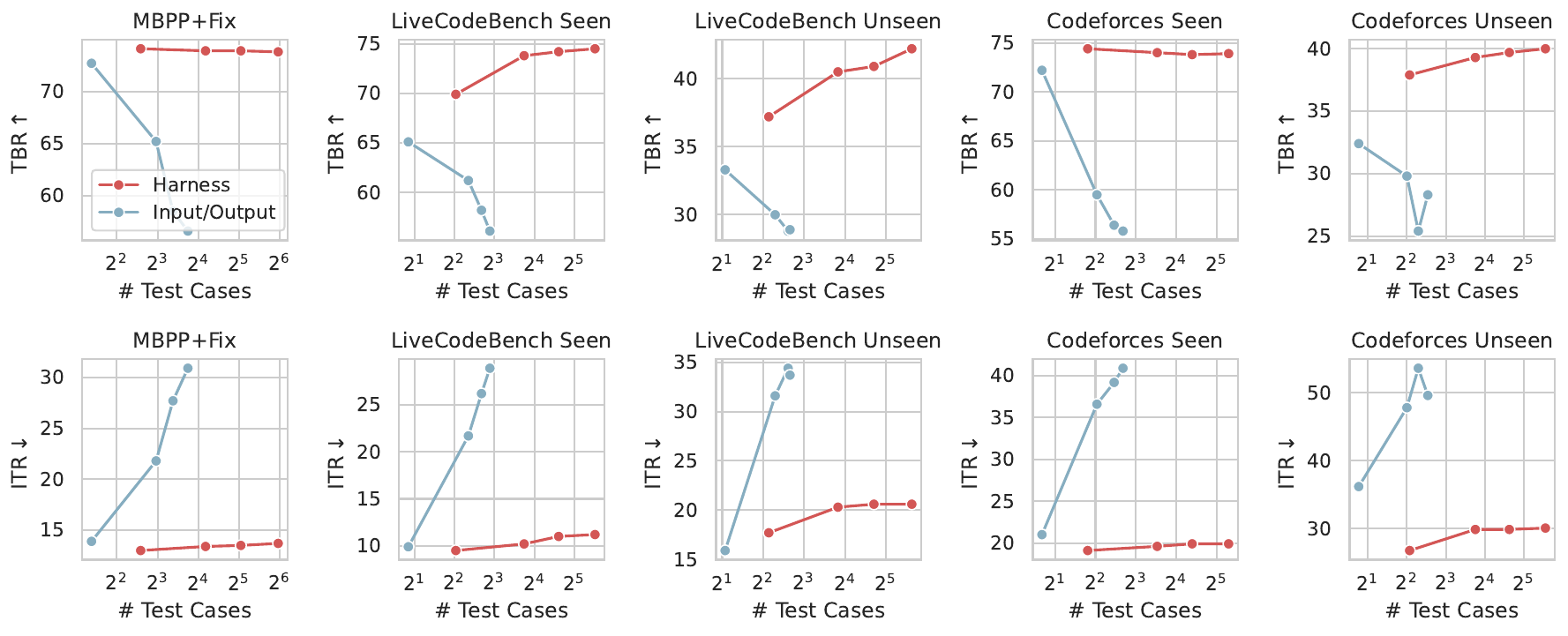}
  \caption{True bug rate (TBR) and invalid test rate (ITR) as the number of test cases increases.}
  \label{fig:num_test}
\end{figure}

\paragraph{Scaling Number of Test Cases.}
In the experiments above, we have limited each response to at most 20 test cases. We next investigate if we can further improve the performance by increasing the number of test cases in each response. Specifically, we employ different strategies to scale up the number of test cases for baselines and our method. For the baseline that generates input-output pairs, we directly change the instruction to the LLM to ask it to generate more test cases. For our method, since many input generators use random functions to generate inputs, we simply run the input generators multiple times with different random seeds to get more test inputs. Figure \ref{fig:num_test} shows the performance of the RL-trained models with respect to the number of test cases. As can be observed, when generating more test cases for the baseline method, the percentage of correctly identified bugs (TBR) drops significantly, and the amount of invalid tests (ITR) quickly increases, leading to a much worse performance. The observation confirms the limitations of hardcoded input-output pairs, since the probability of getting all test cases correct decays exponentially when the number of test cases increases. On the contrary, for our method, TBR consistently increases for three datasets and maintains the original value for the other two datasets, and ITR also demonstrates only a marginal increase. The results highlight two benefits of programmatic input generation and output verification: \ding{182} The input generator can \textit{easily generate more inputs to increase the test coverage}; and \ding{183} The same output verifier can be reused for different inputs \textit{without sacrificing the accuracy.}

\input{tables/test-time-scale}
\paragraph{Using Feedback for Test-time Scaling.}
Given the superior bug-finding performance of our model, we now explore its application to improve code generation via test-time scaling. Specifically, given a coding problem, we sample 8 candidate programs from an LLM and use the test case generator to generate test cases for each program. We collect all generated test cases for the same problem and run them against each candidate program. The program that passes the most test cases is selected as the final program.
Table \ref{tab:test-time-scale} shows the results on 341 problems of \textsc{LiveCodeBench} with three code generators. As can be observed, scaling with both test case generators significantly improves the performance of the original LLM (original pass@1). Furthermore, our model with test harnesses outperforms the input-output testing, demonstrating its superior performance in judging code correctness. The results also confirm that our model's improvements on finding bugs can be \textbf{translated into improved code generation}.

\input{tables/diversity}

\subsection{Additional Analyses}

\paragraph{Diversity of Test Cases.}
By programmatically generating inputs, our model can potentially generate diverse inputs that would be difficult to synthesize with hardcoding. We now verify this by comparing the diversity of inputs generated by the baseline and our models. Specifically, we analyze the test cases generated for the programs of \texttt{Qwen3-32B} in Table \ref{tab:test-time-scale}. We evaluate a subset of 214 problems that take \texttt{stdin} as inputs. For fair comparison, we randomly downsample the generated test cases so that the two models have the same number of test cases. We then calculate three metrics for the two lists of inputs:
\ding{172} Unique ratio: We calculate \e{\frac{\#\text{ of unique inputs}}{\#\text{ of total inputs}}}, where equality is defined by string matching. \ding{173} Length range: We calculate \e{\log(\mathrm{max\_length} + 1) - \log(\mathrm{min\_length} + 1)}, where \e{\mathrm{max\_length}} and \e{\mathrm{min\_length}} are the maximum and minimum lengths of the inputs. \ding{174} Length std: We calculate the standard deviation of the log of each input length.
For all metrics, we compute the value for each individual problem and take the average over all problems. Results in Table \ref{tab:input_diversity} show that our method generates more diverse inputs and inputs with various lengths than the baseline.

\begin{figure}
\centering
\begin{minipage}[t]{0.45\textwidth}
    \vspace{1pt}\centering
    \includegraphics[width=\linewidth]{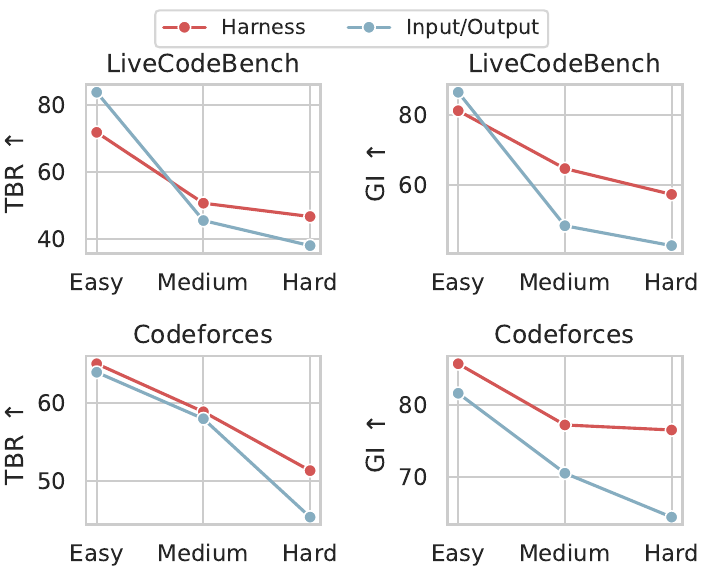}
    \caption{Performance across difficulty levels.}
    \label{fig:perf_breakdown}
\end{minipage}
\hspace{0.01\textwidth}
\begin{minipage}[t]{0.49\textwidth}
    \vspace{1pt}\centering
    \includegraphics[width=\linewidth]{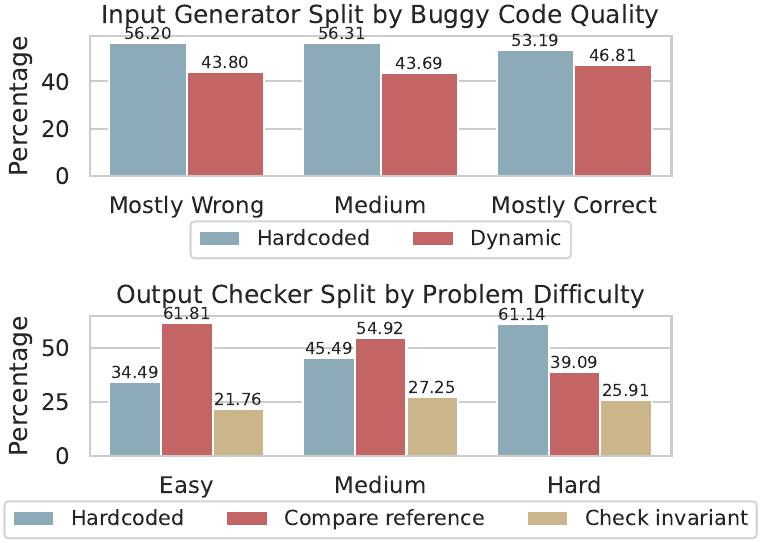}
    \caption{Distribution of testing strategies.}
    \label{fig:input_output}
\end{minipage}
\vspace{-2mm}
\end{figure}

\paragraph{Performance across Difficulty Levels.}
Section \ref{subsec:main-result} reports aggregated performance across all problems in a dataset. We next investigate if the improvement of our method is consistent across problems with different difficulty levels. Figure \ref{fig:perf_breakdown} shows the detailed performance breakdown of the baseline and our method. Specifically, on \textsc{LiveCodeBench}, we use the original difficulty categories. On \textsc{Codeforces}, we split problems based on their ratings (\textsc{Hard} corresponds to problems with ratings greater than 2400 and \textsc{Medium} corresponds to problems with ratings greater than 1800). As can be observed, while the performance of both methods degrades when problems become harder, our method better maintains the performance compared to the baseline. The results indicate that test harnesses can better generalize to difficult problems, verifying our motivation that input-output testing is limited for complex problems.

\paragraph{Distribution of Testing Strategies.}
By programmatically generating inputs and validating outputs, test harnesses allow models to have broader strategies for debugging. For example, we identify two main ways models use to generate inputs, which are explicitly emphasized in our SFT data: \ding{172} \textit{Hardcoded}: models return a list of hardcoded inputs. \ding{173} \textit{Dynamic}: models dynamically generates inputs with code (\emph{e.g.,} randomized inputs through random functions). Similarly, we identify three ways models employ to validate a captured output: \ding{172} \textit{Hardcoded}: models compare the output with a hardcoded expected output. \ding{173} \textit{Compare reference}: models implement a reference solution (\emph{e.g.,} a brute-force solution) and compare the output with that obtained from the reference solution. \ding{174} \textit{Check invariant}: models check if the output satisfies specific invariants such as the length and range.

We prompt \texttt{Qwen3-32B} to classify the strategies used in each response (details in Appendix \ref{append:classify-strategy}). Figure \ref{fig:input_output} shows the distributions of the input generator and the output verifier respectively. Specifically, we report input generator strategies for buggy programs that are mostly wrong (pass less than 25\% of test cases), medium (pass 25\% to 75\% of test cases), and mostly correct (pass greater than 75\% of test cases). As can be observed, when the buggy program is mostly wrong and has obvious bugs, the model generates more hardcoded inputs. When the buggy program is more correct and contains bugs that are hard to identify, the model generates more dynamic inputs to increase test coverage.

\begin{figure}
  \centering
  \includegraphics[width=0.95\linewidth]{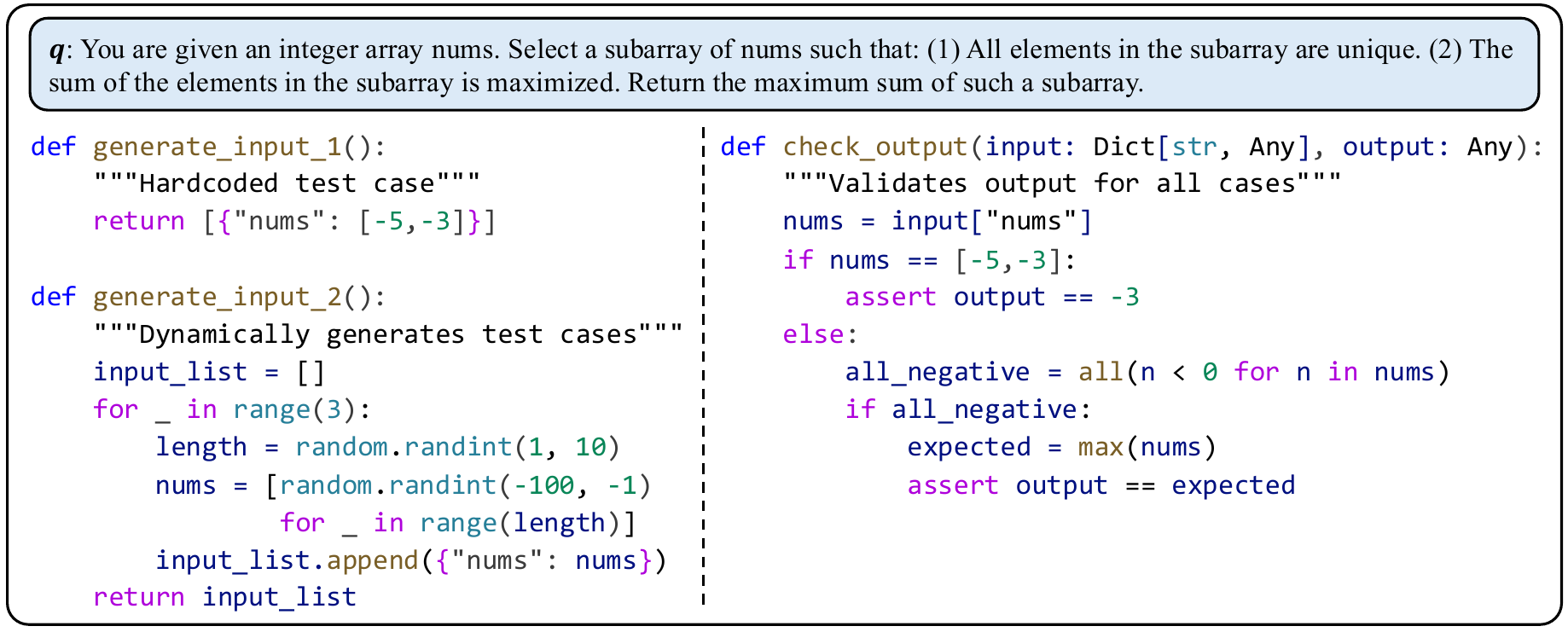}
    \caption{A sample output using a combination of strategies for input generators and output verifier.}
    \label{fig:sample}
\end{figure}

Similarly, when the problem is easy, the model more often implements a reference solution for validation;\footnote{An output verifier can use a combination of strategies, so the numbers do not add up to 100.} and when the problem becomes difficult, the model hardcodes more expected outputs. The observations demonstrate that the model can adapt its testing strategies to specific problems. Figure \ref{fig:sample} shows an example where the model combines multiple strategies for output validation.

%% file: tables/main_result.tex
\begin{table}[t]
\begin{center}
\caption{Performance on finding bugs (average of 8 runs). $^*$: The model and training set are not released, so we compare with the number reported in the original paper. Note that the results of \texttt{Qwen3-32B} come from the original model without any fine-tuning.}
\label{tab:main-result}
\resizebox{0.95\textwidth}{!}{
\begin{tabular}{l|ccc|ccc|ccc}
     \toprule \midrule
     & \multicolumn{3}{c|}{\textbf{\textsc{MBPP+Fix (Hard)}}} & \multicolumn{3}{c|}{\textbf{\textsc{LiveCodeBench}}} & \multicolumn{3}{c}{\textbf{\textsc{Codeforces}}} \\
     & GI $\uparrow$ & ITR $\downarrow$ & TBR $\uparrow$ & GI $\uparrow$ & ITR $\downarrow$ & TBR $\uparrow$ & GI $\uparrow$ & ITR $\downarrow$ & TBR $\uparrow$ \\
     \midrule
\texttt{UTGen-32B}$^*$ \cite{prasad2025learninggenerateunittests} & 56.1 & 40.8 & 34.7 & -- & -- & -- & -- & -- & -- \\
\texttt{Qwen3-32B (Input/Output)} & 56.4 & 10.1 & 49.3 & 56.7 & 5.1 & 54.8 & 79.9 & 21.6 & 67.1 \\
\texttt{Qwen3-32B (Harness)} & 78.7 & 11.9 & 68.6 & 69.1 & 15.5 & 55.1 & 80.4 & 33.9 & 54.8 \\
\midrule
\multicolumn{10}{c}{\textbf{\texttt{Qwen3-4B}}} \\
\midrule
\texttt{SFT (Input/Output)} & 52.1 & 11.3 & 44.6 & 45.7 & 8.2 & 42.9 & 75.1 & 23.6 & 59.5 \\
\texttt{SFT (Harness)} & 78.1 & 17.7 & 62.9 & 60.4 & 23.7 & 42.1 & 82.5 & 46.9 & 46.0 \\
\midrule
\texttt{RL (Input/Output)} & 82.5 & 13.9 & 72.7 & 68.4 & 9.9 & 65.1 & 89.8 & 21.0 & 72.2 \\
\rowcolor{gray!20}
\texttt{RL (Harness)} & \textbf{84.4} & \textbf{13.0} & \textbf{74.1} & \textbf{79.1} & \textbf{9.5} & \textbf{69.9} & \textbf{91.8} & \textbf{19.1} & \textbf{74.4} \\
\midrule \bottomrule
\end{tabular}
}
\end{center}
\end{table}

%% file: tables/generalization.tex
\begin{wrapfigure}{r}{0.5\textwidth}
\centering
\vspace{-4mm}
\captionof{table}{Generalization to unseen models. The buggy code is sampled from \texttt{Qwen3-14B}, which is not seen during training.}
\label{tab:generalization}
\resizebox{\linewidth}{!}{
\begin{tabular}{l|ccc|ccc}
     \toprule \midrule
     & \multicolumn{3}{c|}{\textbf{\textsc{LiveCodeBench}}} & \multicolumn{3}{c}{\textbf{\textsc{Codeforces}}} \\
     & GI $\uparrow$ & ITR $\downarrow$ & TBR $\uparrow$ & GI $\uparrow$ & ITR $\downarrow$ & TBR $\uparrow$ \\
     \midrule
\multicolumn{7}{c}{\textbf{\texttt{Qwen3-32B}}} \\
\midrule
\texttt{I/O} & 25.0 & 8.3 & 23.0 & 43.2 & 20.1 & 31.5 \\
\texttt{Harness} & 36.8 & 20.4 & 22.4 & 61.6 & 36.3 & 32.3 \\
\midrule
\multicolumn{7}{c}{\textbf{\texttt{Qwen3-4B}}} \\
\midrule
\texttt{SFT (I/O)} & 19.4 & 12.4 & 17.3 & 35.7 & 25.4 & 23.1 \\
\texttt{SFT (Har)} & 34.1 & 34.5 & 16.3 & 59.1 & 45.2 & 25.3 \\
\midrule
\texttt{RL (I/O)} & 37.0 & \textbf{15.9} & 33.3 & 53.0 & 36.2 & 32.4 \\
\rowcolor{gray!20}
\texttt{RL (Har)} & \textbf{51.1} & 17.7 & \textbf{37.2} & \textbf{67.3} & \textbf{26.8} & \textbf{37.9} \\
\midrule \bottomrule
\end{tabular}
}
\vspace{-2mm}
\end{wrapfigure}

%% file: tables/test-time-scale.tex
\begin{wrapfigure}{r}{0.5\textwidth}
\centering
\vspace{-4mm}
\captionof{table}{Best-of-8 performance on \textsc{LiveCodeBench} where the code is selected based on the execution results of the generated test cases.}
\label{tab:test-time-scale}
\resizebox{\linewidth}{!}{
\begin{tabular}{l|ccc}
     \toprule \midrule
     & \multicolumn{3}{c}{\textbf{Code Generator}} \\
     & \textbf{\texttt{Qwen3-4B}} & \textbf{\texttt{Qwen3-14B}} & \textbf{\texttt{Qwen3-32B}} \\
\midrule
Original pass@1 & 52.60 & 60.23 & 63.53 \\
\texttt{RL (I/O)} & 60.12 & 65.40 & 67.45 \\
\rowcolor{gray!20}
\texttt{RL (Harness)} & \textbf{60.70} & \textbf{66.57} & \textbf{69.50} \\
\midrule \bottomrule
\end{tabular}
}
\end{wrapfigure}

%% file: tables/diversity.tex
\begin{wrapfigure}{r}{0.5\textwidth}
\centering
\vspace{-4mm}
\captionof{table}{Diversity of inputs for test cases generated by the two models.}
\label{tab:input_diversity}
\vspace{-1mm}
\resizebox{\linewidth}{!}{
\begin{tabular}{l|ccc}
     \toprule \midrule
     & \textbf{Unique ratio $\uparrow$} & \textbf{Length range $\uparrow$} & \textbf{Length std $\uparrow$} \\
\midrule
\texttt{RL (I/O)} & 48.6 & 1.00 & 0.31 \\
\rowcolor{gray!20}
\texttt{RL (Har)} & 77.1 & 8.90 & 2.69 \\
\midrule \bottomrule
\end{tabular}
}
\end{wrapfigure}

%% file: section/5_conclusion.tex
\section{Conclusion and Future Works}
We propose \sys, a pipeline for training LLMs for test harness generation. Through two-stage training of SFT followed by RLVR, we demonstrate that \sys outperforms its counterpart that generates input-output pairs. Additional experiments show that \sys exhibits better generalizability and benefits the code generation performance with test-time scaling.

One of the future directions is to explore methods that reduce the reliance on ground-truth programs. Currently, our method requires access to ground-truth programs during training to ensure the model generates valid test cases. In situations where the ground-truth programs are difficult to obtain, future works could explore directions such as using weaker oracles or generalizing models trained on simpler tasks with ground truths to more difficult tasks.

%% file: section/6_addition.tex
\section{Acknowledgements}
The work of Yujian Liu, Jiabao Ji, and Shiyu Chang was partially supported by National Science Foundation (NSF) Grant IIS-2338252, NSF Grant IIS-2207052, and NSF Grant IIS-2302730.

\section{Ethics Statement}
This work aims to enhance the reliability and robustness of AI-generated programs by developing improved methods for testing and debugging. However, while our method shows clear improvements over the baseline, it does not capture all bugs or provide any guarantees on the program's correctness. Our experiments show that some bugs remain hidden and some correct programs may be mistakenly flagged. Therefore, users should remain cautious when interpreting the execution results of our generated test cases. We advise that any use of this system in high-stakes environments be accompanied by additional verification and human oversight.

\section{Reproducibility Statement}
We have taken the necessary steps to ensure the reproducibility of our results. Specifically, Section \ref{subsec:setting} discusses the general experiment settings in our paper. Appendix \ref{append:data} provides the detailed steps to collect the training and evaluation datasets. Finally, Appendix \ref{append:details} lists the implementation details of our method and baselines, including training hyperparameters and evaluation details.

%% file: section/appendix.tex
\section{Dataset Construction}
\label{append:data}

\subsection{Training Data}
\label{append:train-data}
To train LLMs for test case generation, we collect data in the triplets of problem description \e{\bm q}, buggy program \e{f}, and ground-truth program \e{g}. We consider Python programs in this paper. We source such triplets from existing coding datasets, including TACO \cite{li2023tacotopicsalgorithmiccode}, SYNTHETIC-1 \cite{primeintellect_synthetic1_2025}, LeetCode \cite{xia2025leetcodedatasettemporaldatasetrobust}, and Codeforces \cite{matrixstudio_codeforces_python_2025}. These datasets come with the problem description, a ground-truth program, and a list of ground-truth test cases. We use the following three steps to collect data:

\input{tables/train-data}

\noindent
\ding{182} Filter ground-truth programs: We run the given ground-truth program \e{g} on all test cases and only keep problems where \e{g} passes all test cases.

\noindent
\ding{183} Generate buggy programs: We sample candidate programs from \texttt{Qwen2.5-Coder} 1.5B, 3B, 7B \cite{hui2024qwen25codertechnicalreport}, and \texttt{DeepSeek-R1-Distill-Qwen-1.5B} \cite{deepseekai2025deepseekr1incentivizingreasoningcapability}. We sample 8 programs from each model and run the programs on all ground-truth test cases. We only keep programs that pass at least one test case but not all test cases, resulting in partially correct programs. If there are multiple candidates that satisfy the requirement, we use the two that pass the most test cases, which makes it harder to find bugs.

\noindent
\ding{184} Decontamination: We decontaminate training data against all evaluation benchmarks based on the problem description.

We use all collected data for RL training and a subset of data for SFT, ensuring that models see new data during RL training. Table \ref{tab:train-data} shows the statistics of our training set. Specifically, the dataset contains two types of problems: standard input/output problems that read from \texttt{stdin} and return to \texttt{stdout}, as well as functional problems that implement a function in Python. Since the number of functional problems is small, we create two versions for each functional problem, where one contains a few example input-output pairs in the description, and the other does not.

\paragraph{SFT Data.}
To collect SFT data, we use the rejection sampling technique \cite{touvron2023llama2openfoundation}. Specifically, we prompt \texttt{Qwen3-32B} to generate 6 responses for each pair of description and buggy program. Figures~\ref{fig:io-prompt} and \ref{fig:harness-io} show the prompt we use for input-output testing and test harnesses, respectively. Particularly, for harness generation, we encourage the model to use diverse strategies to validate outputs, such as checking specific invariants and comparing with a brute-force solution, which is similar to the strategy used in prior works~\cite{zhang2023algosynthesizingalgorithmicprograms}. We run generated test cases on both ground-truth program \e{g} and buggy program \e{f} and only keep responses where \e{g} passes the test but \e{f} does not. We keep the amount of SFT data the same for input-output testing and harness testing.

\input{tables/eval-data}

\subsection{Evaluation Data}
\label{append:eval-data}
We evaluate on three popular code generation datasets: \textsc{MBPP+} \cite{austin2021programsynthesislargelanguage,liu2023is}, \textsc{LiveCodeBench} \cite{jain2025livecodebench}, and \textsc{Codeforces} \cite{penedo2025codeforces}. Although these datasets are designed for code generation tasks, we convert them into bug-finding tasks following the procedure in Section \ref{append:train-data}.

Specifically, for \textsc{LiveCodeBench}, we use problems from 2024/10 to 2025/4. For \textsc{Codeforces}, we use samples in the test split. For both datasets, we use correct public submissions as the ground-truth program, after rerunning and filtering the submissions on all test cases.

For \textsc{MBPP+}, we directly use the split \textsc{MBPP+Fix (Hard)} in \texttt{UTGen-32B}~\cite{prasad2025learninggenerateunittests}, which is collected similarly to the above procedure. Particularly, we notice the problem descriptions in \textsc{MBPP+} are overly simplified and without clear input specifications (\emph{e.g.,} \textit{`Write a function to find the length of the longest palindromic subsequence in the given string'}, without specifying that the input string should be non-empty). We thus use \texttt{Qwen3-32B} to add an input specification to the problem (detailed prompt in Figure~\ref{fig:input-spec}). To make sure the ground-truth program \e{g} matches the description after modification, we further prompt \texttt{Qwen3-32B} to adapt the original \e{g} to the new description (detailed prompt in Figure~\ref{fig:fix-code}). Finally, we filter the modified ground-truth programs and only keep those that pass the original ground-truth test cases.

Table \ref{tab:eval-data} lists the statistics of all evaluation benchmarks.

\input{tables/num-test-io}

\section{Implementation Details}
\label{append:details}

\subsection{Number of Test Cases}
\label{append:num-test-detail}
For the teacher model and SFT models, we observe that the number of test cases in a response significantly affects the final performance. For example, although we allow models to generate multiple test cases in each response, Tables \ref{tab:num-test-io} and \ref{tab:num-test-harness} show that the performance of \texttt{Qwen3-32B} can vary significantly if we only evaluate the first $k$ test cases. Both methods' performance improves as we evaluate on fewer test cases, especially for input-output-based testing. This confirms the observations in Figure \ref{fig:num_test}, where the performance of input-output testing quickly drops when generating more test cases. Based on these results, for the teacher model and SFT models of input-output testing, we report the performance when $k=1$. For test harnesses, we report the performance when $k=5$.

For the RL models, we observe that the models automatically find a good number of test cases to generate. For instance, the RL trained \texttt{Qwen3-4B} model for input-output testing generates 1.96 test cases in each response on average. Thus, we allow the model itself to determine the number of test cases, and we only restrict the maximum test cases at 20.

\input{tables/hyperparam}

\subsection{Training Hyperparameters}
\label{append:hyperparam}
We run all experiments on 16 NVIDIA H100 GPUs. The RL training for our model takes around 1,500 GPU hours. The RL training for the input-output baseline takes around 1,150 GPU hours. Table \ref{tab:hyperparam} lists the hyperparameters for SFT and RL training. Note that we use the same hyperparameters for all models.

\subsection{Classifying Testing Strategies}
\label{append:classify-strategy}
We prompt \texttt{Qwen3-32B} to identify specific testing strategies used by our model. Specifically, given the generated harness code, we ask the model to identify strategies used in each input generator and output verifier. The detailed prompts are listed in Figures \ref{fig:input-strategy} and \ref{fig:output-strategy}.

\begin{figure}
  \centering
  \includegraphics[width=\linewidth]{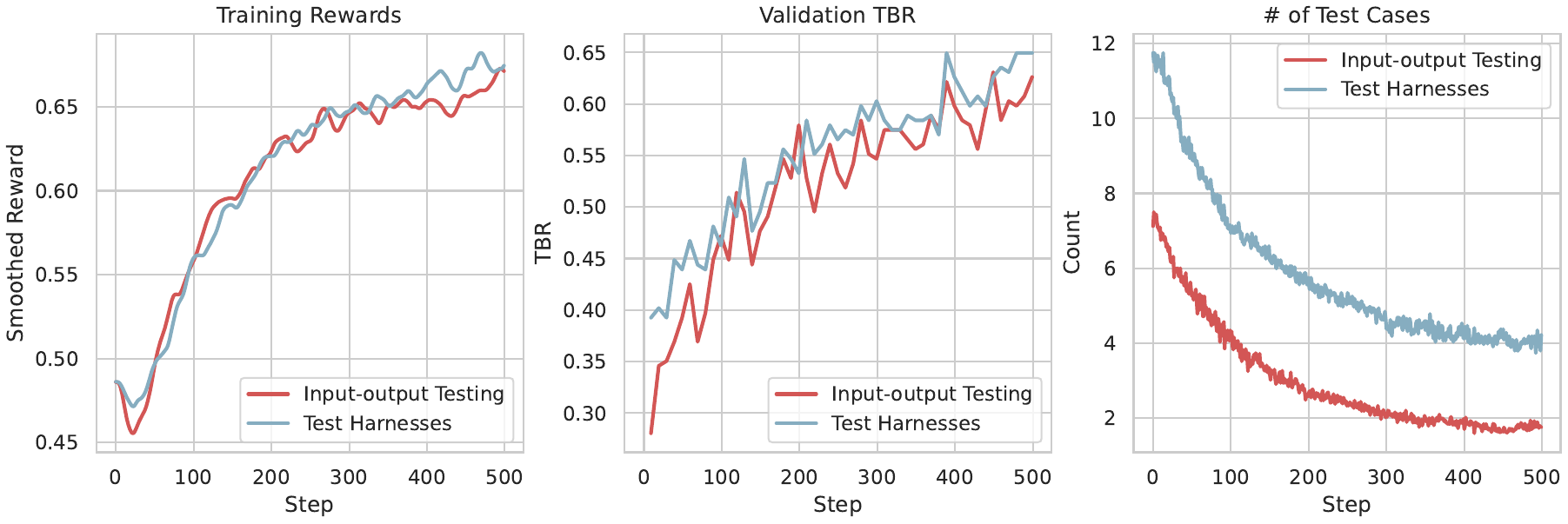}
    \caption{Dynamics of smoothed training rewards, true bug rage (TBR $\uparrow$) on validation set, and the number of generated test cases throughout RL training.}
    \label{fig:training_curve}
\end{figure}

\section{Additional Results}
\subsection{Training Dynamics}
\label{append:dynamics}
Figure~\ref{fig:training_curve} shows the dynamics of RL training for both methods. As can be observed, our method consistently achieves a higher TBR on the validation set than the baseline. Moreover, both methods generate fewer test cases as the training progresses, approaching the optimal number in Section~\ref{append:num-test-detail}. This indicates that the models are learning the best number of test cases for generation.

\subsection{Results on Llama}
\label{append:llama-res}
Table \ref{tab:llama-result} shows the performance when training on \texttt{Llama3.2-3B} model. As can be observed, our model for test harnesses achieves comparable performance with input-output testing on the \textsc{Seen} version of the datasets. However, it significantly outperforms the input-output testing when evaluated on the \textsc{Unseen} version, \emph{e.g.,} a relative improvement over 110\% in TBR on \textsc{LiveCodeBench}. The results indicate that input-output testing has the risk of overfitting to a particular distribution of bugs, whereas test harnesses have better generalizability.

\input{tables/llama-res}
\input{tables/controlled_bug_finding}

\subsection{Performance under Controlled Number of Test Cases}
\label{append:control_test}
Our experiments in Section~\ref{subsec:main-result} demonstrate that different methods should generate different numbers of test cases for the best performance. Particularly, for input-output testing, models usually have better performance when generating fewer test cases, since more test cases lead to a higher probability that one of the test cases is wrong. By contrast, for test harnesses, performance can be further improved by scaling up the number of test cases, which increases the test coverage without sacrificing accuracy. Nevertheless, in the following section, we also report the performance when controlling the number of test cases.

Specifically, we repeat the experiments in Tables~\ref{tab:main-result} and~\ref{tab:generalization} but instruct the input-output testing model to only generate a single test case in each response. For test harnesses, we also prompt the model to generate a single input generator. Then, to get more test cases, we sample multiple responses for input-output testing and run the same input generator multiple times with different random seeds for test harnesses. Table~\ref{tab:controlled-bug-result} presents the performance under this controlled setting. As can be observed, our method consistently outperforms input-output testing, and the gap becomes larger when increasing the number of test cases. These results demonstrate the consistent improvements of the proposed test harness.

\input{tables/controlled_test_time_scale}

Moreover, we also rerun the test-time scaling experiment in Table~\ref{tab:test-time-scale}. Here, we further restrict each input generator to have a single test input, thus ensuring the two methods have the same number of test cases for each candidate program. For both methods, we sample 5 responses for each candidate program to obtain more test cases. The results in Table~\ref{tab:controlled-test-time-scale} show that our method surpasses the baseline in most settings. Particularly, it significantly outperforms the baseline when the strongest \texttt{Qwen3-32B} is used as the code generator, demonstrating the superior generalizability and potential weak-to-strong generalization of our method.

\clearpage

\input{macro}

\begin{tcolorbox}[breakable, enhanced]
\begin{lstlisting}[style=courierStyle]
Given a problem statement and a Python program that aims to solve it, your task is to **write test cases** that uncover any potential bugs.

### **Task Overview**

You should output a JSON object that contains a list of test cases for the provided program. Each test case should include:
1. **input_str**: The exact text to feed into stdin.
2. **expected_output**: The exact text the program should print.

We will run each test by feeding `input_str` into the program and comparing its stdout against `expected_output`.

### **Required Format**

```json
[
  {
    "input_str": "input 1",
    "expected_output": "output 1"
  },
  {
    "input_str": "input 2",
    "expected_output": "output 2"
  }
  // ... up to 20 test cases total
]
```

### **Constraints**

* Generate **1-20** test cases.
* Don't include comments or extra fields in the JSON.
* Each input_str and expected_output must be a valid JSON string.

The problem is as follows:
{description}

And the program is as follows:
```python
{target_code}
```
\end{lstlisting}
\end{tcolorbox}
\captionof{figure}{Prompt used for input-output-based testing. Note that this prompt assumes the program reads input from stdin.}
\label{fig:io-prompt}

\newpage

\begin{tcolorbox}[breakable, enhanced]
\begin{lstlisting}[style=courierStyle]
Given a problem statement and a Python program that aims to solve it, your task is to **write a test harness** that uncovers any potential bugs.

### **Task Overview**

You will deliver **a single** code block to define functions that can be run by our framework to generate inputs, run the program, and validate its outputs.
Consider two categories of test cases:
- **Hardcoded cases**: Manually crafted input-output pairs that expose known or likely bugs.
- **Dynamic cases**: Programmatically generated inputs that stress-test the implementation (e.g., randomized, combinatorial, large or edge-case inputs).

### **Required Functions**

```python
from typing import List

def generate_input_1() -> List[str]:
    """
    Return between 1 and 4 valid input strings, each a complete stdin payload for the target program.
    Consider the following strategies:
      - Manually craft inputs that expose bugs.
      - Dynamically generate randomized, combinatorial, large, or edge-case inputs for stress testing.
    """
    # Your code here
    return input_list

def generate_input_2() -> List[str]:
    """
    Another function to return between 1 and 4 valid input strings.
    Employ a different strategy than previous input generation functions.
    """
    # Your code here
    return input_list

# You may add up to 3 more functions named generate_input_3(), generate_input_4(), etc.

def check_output(generated_input: str, captured_output: str) -> None:
    """
    Validate the output for a single generated input.
    Inputs:
        - generated_input: The input string passed to the target program.
        - captured_output: The exact stdout produced by the target program.
    
    Hints: When exact outputs are hard to predict, avoid asserting them. Instead, consider:
      - Check key properties or invariants, e.g., output is sorted, has correct length, matches a pattern, has correct value ranges, etc.
      - Compare against a simple brute-force implementation
    """
    # Your code here
```

### **Execution Flow**

1. The framework calls generate input functions to obtain a list of test strings.
2. For each string:
   * It runs the target program with that string on stdin.
   * Captures stdout into `captured_output`.
   * Calls `check_output(generated_input, captured_output)`.
3. If any assertion fails, the test suite reports an error.

### **Constraints**

* Provide one contiguous block of Python code that defines all required/optional functions. Do not invoke the functions yourself-only define them.
* Define up to 5 input generation functions, each returning between 1 and 4 inputs.
* The dynamic input functions must employ diverse strategies to generate inputs. Avoid generating inputs with the same logic or from the same distribution.
* Runtime limit per check_output call: 5 seconds.

The problem is as follows:
{description}

And the program is as follows:
```python
{target_code}
```
\end{lstlisting}
\end{tcolorbox}
\captionof{figure}{Prompt used for test harnesses generation. Note that this prompt assumes the program reads input from stdin.}
\label{fig:harness-io}

\newpage

\begin{tcolorbox}[breakable, enhanced]
\begin{lstlisting}[style=courierStyle]
Given the following coding problem and a corresponding solution, improve the problem description by adding input specifications. Include details such as:
- Valid input types (e.g. "integer", "string", "list of floats").
- Reasonable value ranges (e.g. "0 <= n <= 1000").
- Format constraints (e.g. "no empty strings", "no null/None values").

Do not change the original requirements or add example cases, just append the specifications.

Problem:
{problem}

Code:
```python
{code}
```
\end{lstlisting}
\end{tcolorbox}
\captionof{figure}{Prompt used for adding input specifications on \textsc{MBPP+}.}
\label{fig:input-spec}

\begin{tcolorbox}[breakable, enhanced]
\begin{lstlisting}[style=courierStyle]
Given the following coding problem and a corresponding solution, decide whether the solution contains a bug or not. If yes, rewrite the code to fix the bug. Remember to look for edge cases where the code fails to handle.

Problem:
{problem}

Code:
```python
{code}
```

Output your answer in the following format:
```python
fixed_code
```
where fixed_code is the rewritten code that fixes the bug. If the code is correct, just return the original code without any changes.
\end{lstlisting}
\end{tcolorbox}
\captionof{figure}{Prompt used for adapting the ground-truth programs to the new descriptions on \textsc{MBPP+}.}
\label{fig:fix-code}

\newpage

\begin{tcolorbox}[breakable, enhanced]
\begin{lstlisting}[style=courierStyle]
Given the following code snippet for a test harness, determine the strategy used in each `generate_input` function.

Code:
```python
{code}
```

Select from the following options:
- hardcoded: the function returns hardcoded inputs.
- dynamic: the function generates inputs dynamically, e.g., random sampling, or combinatorial generation.

Think about the code step by step and then output your final answer in the following format:
```json
<used strategies>
```
where <used strategies> is a list of the strategies used in each function.

Notes:
- The list should have the same length as the number of `generate_input` functions in the code.
- If a function uses a combination of the above strategies, select the dominant strategy.
\end{lstlisting}
\end{tcolorbox}
\captionof{figure}{Prompt used for identifying strategies in input generators.}
\label{fig:input-strategy}

\begin{tcolorbox}[breakable, enhanced]
\begin{lstlisting}[style=courierStyle]
Given the following code snippet for a test harness, determine the strategies used in the `check_output` function.

Code:
```python
{code}
```

Select from the following options:
- reference implementation: the function compares the output with a reference implementation, e.g., a brute-force solution, or a correct implementation.
- invariant checking: the function checks whether the output satisfies certain invariants or properties, e.g., whether the output is sorted, or whether the output has valid types and lengths.
- hardcoded: the function compares the output with hardcoded expected outputs.

Think about the code step by step and then output your final answer in the following format:
```json
<used strategies>
```
where <used strategies> is a list of the strategies used in the function.

Notes:
- If the function uses a combination of the above strategies, return a list containing all the strategies used, e.g., ["reference implementation", "invariant checking"].
- If the function does not contain any of the above strategies, return an empty list [].
\end{lstlisting}
\end{tcolorbox}
\captionof{figure}{Prompt used for identifying strategies in the output verifier.}
\label{fig:output-strategy}

%% file: tables/train-data.tex
\begin{wrapfigure}{r}{0.45\textwidth}
\centering
\vspace{-4mm}
\captionof{table}{Statistics of our training data.}
\label{tab:train-data}
\resizebox{\linewidth}{!}{
\begin{tabular}{ll}
\toprule \midrule
 & \textbf{Statistic} \\
\midrule
\# triplets for RL & 12,043 \\
\# unique problems for RL & 7,748 \\
\# triplets for SFT & 6,805 \\
\# unique problems for SFT & 4,383 \\
\# responses for SFT & 15,619 \\
\midrule \bottomrule
\end{tabular}
}
\end{wrapfigure}

%% file: tables/eval-data.tex
\begin{wrapfigure}{r}{0.45\textwidth}
\centering
\vspace{-4mm}
\captionof{table}{Statistics of evaluation datasets.}
\label{tab:eval-data}
\resizebox{\linewidth}{!}{
\begin{tabular}{ll}
\toprule \midrule
 & \textbf{\# data} \\
\midrule
\textsc{MBPP+Fix (Hard)} & 141 \\
\textsc{LiveCodeBench Seen} & 76 \\
\textsc{LiveCodeBench Unseen} & 93 \\
\textsc{Codeforces Seen} & 100 \\
\textsc{Codeforces Unseen} & 84 \\
\midrule \bottomrule
\end{tabular}
}
\end{wrapfigure}

%% file: tables/num-test-io.tex
\begin{table}[t]
\centering
\resizebox{\textwidth}{!}{
\begin{minipage}[t]{0.5\textwidth}
\centering
\caption{True bug rate (higher is better) of input-output-based testing with \texttt{Qwen3-32B} when only evaluating the first $k$ generated test cases. We use the \textsc{Seen} version of \textsc{LiveCodeBench} and \textsc{Codeforces}.}
\label{tab:num-test-io}
\resizebox{\linewidth}{!}{
\begin{tabular}{lccc}
\toprule \midrule
 & \textsc{MBPP+} & \textsc{LiveCodeBench} & \textsc{Codeforces} \\
\midrule
$k=1$ & 49.3 & 54.8 & 67.1 \\
$k=3$ & 59.0 & 54.6 & 53.2 \\
$k=5$ & 57.4 & 53.6 & 42.5 \\
$k=10$ & 54.6 & 51.3 & 39.5 \\
\midrule \bottomrule
\end{tabular}
}
\end{minipage}

\hspace{0.01\textwidth}
\begin{minipage}[t]{0.5\textwidth}
\centering
\caption{True bug rate (higher is better) of test harnesses with \texttt{Qwen3-32B} when only evaluating the first $k$ generated test cases. We use the \textsc{Seen} version of \textsc{LiveCodeBench} and \textsc{Codeforces}.}
\label{tab:num-test-harness}
\resizebox{\linewidth}{!}{
\begin{tabular}{lccc}
\toprule \midrule
 & \textsc{MBPP+} & \textsc{LiveCodeBench} & \textsc{Codeforces} \\
\midrule
$k=3$ & 66.6 & 48.5 & 57.9 \\
$k=5$ & 68.6 & 55.1 & 54.8 \\
$k=10$ & 67.7 & 53.8 & 48.6 \\
$k=20$ & 67.3 & 53.3 & 44.2 \\
\midrule \bottomrule
\end{tabular}
}
\end{minipage}
}
\end{table}

%% file: tables/hyperparam.tex
\begin{wrapfigure}{r}{0.32\textwidth}
\centering
\vspace{-17mm}
\captionof{table}{Training hyperparameters. The same hyperparameters are used for all models.}
    \label{tab:hyperparam}
    \resizebox{\linewidth}{!}{
    \begin{tabular}{lc}
    \toprule \midrule
     \multicolumn{2}{c}{\textbf{SFT Training}} \\
     \midrule
     \# Epochs & 15 \\
     Batch size & 96 \\
     Learning rate & $1e{-5}$ \\
     LR scheduler & cosine \\
     \midrule
     \multicolumn{2}{c}{\textbf{RL Training}} \\
     \midrule
     \# Steps & 500 \\
     Batch size & 128 \\
     \# Rollouts per question & 8 \\
     Learning rate & $1e{-6}$ \\
     LR scheduler & None \\
     Max response length & 16,384 \\
     \midrule
     \bottomrule
    \end{tabular}
    }
\vspace{-14mm}
\end{wrapfigure}

%% file: tables/llama-res.tex
\begin{table}
\begin{center}
\resizebox{\textwidth}{!}{
\begin{tabular}{l|ccc|ccc|ccc|ccc|ccc}
     \toprule \midrule
     & \multicolumn{3}{c|}{\textbf{\textsc{MBPP+Fix (Hard)}}} & \multicolumn{3}{c|}{\textbf{\textsc{LCB Seen}}} & \multicolumn{3}{c|}{\textbf{\textsc{CF Seen}}} & \multicolumn{3}{c|}{\textbf{\textsc{LCB Unseen}}} & \multicolumn{3}{c}{\textbf{\textsc{CF Unseen}}} \\
     & GI $\uparrow$ & ITR $\downarrow$ & TBR $\uparrow$ & GI $\uparrow$ & ITR $\downarrow$ & TBR $\uparrow$ & GI $\uparrow$ & ITR $\downarrow$ & TBR $\uparrow$ & GI $\uparrow$ & ITR $\downarrow$ & TBR $\uparrow$ & GI $\uparrow$ & ITR $\downarrow$ & TBR $\uparrow$ \\
     \midrule
\texttt{RL (I/O)} & 71.3 & 37.3 & \textbf{45.3} & 47.0 & 42.3 & 30.3 & 67.6 & 53.9 & \textbf{32.8} & 21.0 & 61.8 & 8.3 & 37.4 & 66.2 & 10.1 \\
\rowcolor{gray!20}
\texttt{RL (Har)} & \textbf{77.9} & 37.3 & 42.6 & \textbf{76.5} & \textbf{33.9} & \textbf{31.4} & \textbf{81.4} & \textbf{43.8} & 29.9 & \textbf{59.9} & \textbf{39.9} & \textbf{17.7} & \textbf{73.4} & \textbf{43.3} & \textbf{21.6} \\
\midrule \bottomrule
\end{tabular}
}
\caption{Performance of \texttt{Llama3.2-3B} on finding bugs (average of 8 runs). \texttt{I/O}: input-output testing. \texttt{Har}: test harnesses.}
\label{tab:llama-result}
\end{center}
\end{table}

%% file: tables/controlled_bug_finding.tex
\begin{table}[t]
\begin{center}
\resizebox{\textwidth}{!}{
\begin{tabular}{l|ccc|ccc|ccc|ccc|ccc}
     \toprule \midrule
     & \multicolumn{3}{c|}{\textbf{\textsc{MBPP+Fix (Hard)}}} & \multicolumn{3}{c|}{\textbf{\textsc{LCB Seen}}} & \multicolumn{3}{c|}{\textbf{\textsc{CF Seen}}} & \multicolumn{3}{c|}{\textbf{\textsc{LCB Unseen}}} & \multicolumn{3}{c}{\textbf{\textsc{CF Unseen}}} \\
     & GI $\uparrow$ & ITR $\downarrow$ & TBR $\uparrow$ & GI $\uparrow$ & ITR $\downarrow$ & TBR $\uparrow$ & GI $\uparrow$ & ITR $\downarrow$ & TBR $\uparrow$ & GI $\uparrow$ & ITR $\downarrow$ & TBR $\uparrow$ & GI $\uparrow$ & ITR $\downarrow$ & TBR $\uparrow$ \\
     \midrule
& \multicolumn{15}{c}{\textbf{Repeat Once}} \\
\midrule
\texttt{RL (I/O)} & 80.5 & 11.6 & 73.8 & 67.4 & 9.7 & 65.1          & 87.8 & 16.2 & 75.6 & 36.0 & 18.7 & 33.9 & 49.1 & 30.7 & 34.2 \\
\rowcolor{gray!20}
\texttt{RL (Har)} & 83.6 & 12.1 & \textbf{74.9} & 75.0 & 7.9 & \textbf{68.4} & 90.2 & 16.1 & \textbf{76.9} & 43.3 & 13.6 & \textbf{37.1} & 58.3 & 21.4 & \textbf{34.5} \\
\midrule
& \multicolumn{15}{c}{\textbf{Repeat 5 Times}} \\
\midrule
\texttt{RL (I/O)} & 86.7 & 22.3 & 69.5 & 79.9 & 23.7 & 63.2 & 97.2 & 34.8 & 63.8 & 54.8 & 46.5 & 31.7 & 75.0 & 56.2 & 31.8 \\
\rowcolor{gray!20}
\texttt{RL (Har)} & 83.8 & 12.3 & \textbf{75.0} & 75.0 & 7.9 & \textbf{68.4} & 90.8 & 16.1 & \textbf{77.4} & 49.5 & 14.5 & \textbf{40.9} & 61.9 & 22.6 & \textbf{34.5} \\
\midrule
& \multicolumn{15}{c}{\textbf{Repeat 20 Times}} \\
\midrule
\texttt{RL (I/O)} & 88.7 & 34.0 & 60.3 & 86.8 & 42.1 & 51.3 & 100.0 & 56.0 & 44.0 & 64.5 & 72.0 & 18.3 & 86.9 & 79.8 & 16.7 \\
\rowcolor{gray!20}
\texttt{RL (Har)} & 83.5 & 12.5 & \textbf{74.6} & 77.6 & 7.9 & \textbf{71.1} & 90.8 & 16.4 & \textbf{77.1} & 52.2 & 14.9 & \textbf{42.5} & 63.1 & 23.8 & \textbf{35.7} \\
\midrule \bottomrule
\end{tabular}
}
\caption{Bug-finding performance of \texttt{Qwen3-4B} when controlling the number of test cases. \texttt{I/O}: input-output testing. \texttt{Har}: test harnesses.}
\label{tab:controlled-bug-result}
\end{center}
\end{table}

%% file: tables/controlled_test_time_scale.tex
\begin{wrapfigure}{r}{0.5\textwidth}
\centering
\vspace{-4mm}
\captionof{table}{Best-of-8 performance on \textsc{LiveCodeBench} where the code is selected based on the execution results of the generated test cases.}
\label{tab:controlled-test-time-scale}
\resizebox{\linewidth}{!}{
\begin{tabular}{l|ccc}
     \toprule \midrule
     & \multicolumn{3}{c}{\textbf{Code Generator}} \\
     & \textbf{\texttt{Qwen3-4B}} & \textbf{\texttt{Qwen3-14B}} & \textbf{\texttt{Qwen3-32B}} \\
\midrule
Original pass@1 & 52.60 & 60.23 & 63.53 \\
\midrule
& \multicolumn{3}{c}{\textbf{1 test case per program}} \\
\midrule
\texttt{RL (I/O)} & 60.41 & \textbf{65.10} & 65.98 \\
\rowcolor{gray!20}
\texttt{RL (Harness)} & \textbf{60.70} & 64.81 & \textbf{68.33} \\
\midrule
& \multicolumn{3}{c}{\textbf{5 test cases per program}} \\
\midrule
\texttt{RL (I/O)} & \textbf{61.88} & 67.16 & 68.04 \\
\rowcolor{gray!20}
\texttt{RL (Harness)} & 61.00 & \textbf{67.74} & \textbf{72.14} \\
\midrule \bottomrule
\end{tabular}
}
\end{wrapfigure}

%% file: macro.tex
\makeatletter
\lst@InstallKeywords k{attributes}{attributestyle}\slshape{attributestyle}{}ld
\makeatother

\lstdefinestyle{courierStyle}{
    basicstyle=\fontsize{8}{9}\fontfamily{pcr}\selectfont,
    showstringspaces=false,
    breaklines=true,
    breakatwhitespace=false,
    breakindent=0pt,
    keepspaces=false,
    showspaces=false,   
    escapeinside={(*@}{@*)}
}

%% file: iclr2026_conference.bbl
\begin{thebibliography}{53}
\providecommand{\natexlab}[1]{#1}
\providecommand{\url}[1]{\texttt{#1}}
\expandafter\ifx\csname urlstyle\endcsname\relax
  \providecommand{\doi}[1]{doi: #1}\else
  \providecommand{\doi}{doi: \begingroup \urlstyle{rm}\Url}\fi

\bibitem[Austin et~al.(2021)Austin, Odena, Nye, Bosma, Michalewski, Dohan, Jiang, Cai, Terry, Le, and Sutton]{austin2021programsynthesislargelanguage}
Jacob Austin, Augustus Odena, Maxwell Nye, Maarten Bosma, Henryk Michalewski, David Dohan, Ellen Jiang, Carrie Cai, Michael Terry, Quoc Le, and Charles Sutton.
\newblock Program synthesis with large language models, 2021.

\bibitem[Cao et~al.(2025)Cao, Chen, Quan, Zhang, Wang, Dong, Feng, He, Huang, Li, Tan, Tang, Tang, Wu, Xiao, Zheng, Zhou, Zhu, Huang, Xie, and He]{cao2025llmsgeneratereliabletest}
Yuhan Cao, Zian Chen, Kun Quan, Ziliang Zhang, Yu~Wang, Xiaoning Dong, Yeqi Feng, Guanzhong He, Jingcheng Huang, Jianhao Li, Yixuan Tan, Jiafu Tang, Yilin Tang, Junlei Wu, Qianyu Xiao, Can Zheng, Shouchen Zhou, Yuxiang Zhu, Yiming Huang, Tian Xie, and Tianxing He.
\newblock Can llms generate reliable test case generators? a study on competition-level programming problems, 2025.

\bibitem[Chen et~al.(2022)Chen, Zhang, Nguyen, Zan, Lin, Lou, and Chen]{chen2022codetcodegenerationgenerated}
Bei Chen, Fengji Zhang, Anh Nguyen, Daoguang Zan, Zeqi Lin, Jian-Guang Lou, and Weizhu Chen.
\newblock Codet: Code generation with generated tests, 2022.

\bibitem[Chen et~al.(2021)Chen, Tworek, Jun, Yuan, de~Oliveira~Pinto, Kaplan, Edwards, Burda, Joseph, Brockman, Ray, Puri, Krueger, Petrov, Khlaaf, Sastry, Mishkin, Chan, Gray, Ryder, Pavlov, Power, Kaiser, Bavarian, Winter, Tillet, Such, Cummings, Plappert, Chantzis, Barnes, Herbert-Voss, Guss, Nichol, Paino, Tezak, Tang, Babuschkin, Balaji, Jain, Saunders, Hesse, Carr, Leike, Achiam, Misra, Morikawa, Radford, Knight, Brundage, Murati, Mayer, Welinder, McGrew, Amodei, McCandlish, Sutskever, and Zaremba]{chen2021evaluatinglargelanguagemodels}
Mark Chen, Jerry Tworek, Heewoo Jun, Qiming Yuan, Henrique~Ponde de~Oliveira~Pinto, Jared Kaplan, Harri Edwards, Yuri Burda, Nicholas Joseph, Greg Brockman, Alex Ray, Raul Puri, Gretchen Krueger, Michael Petrov, Heidy Khlaaf, Girish Sastry, Pamela Mishkin, Brooke Chan, Scott Gray, Nick Ryder, Mikhail Pavlov, Alethea Power, Lukasz Kaiser, Mohammad Bavarian, Clemens Winter, Philippe Tillet, Felipe~Petroski Such, Dave Cummings, Matthias Plappert, Fotios Chantzis, Elizabeth Barnes, Ariel Herbert-Voss, William~Hebgen Guss, Alex Nichol, Alex Paino, Nikolas Tezak, Jie Tang, Igor Babuschkin, Suchir Balaji, Shantanu Jain, William Saunders, Christopher Hesse, Andrew~N. Carr, Jan Leike, Josh Achiam, Vedant Misra, Evan Morikawa, Alec Radford, Matthew Knight, Miles Brundage, Mira Murati, Katie Mayer, Peter Welinder, Bob McGrew, Dario Amodei, Sam McCandlish, Ilya Sutskever, and Wojciech Zaremba.
\newblock Evaluating large language models trained on code, 2021.

\bibitem[Chen et~al.(2024{\natexlab{a}})Chen, Lin, Sch{\"a}rli, and Zhou]{chen2024teaching}
Xinyun Chen, Maxwell Lin, Nathanael Sch{\"a}rli, and Denny Zhou.
\newblock Teaching large language models to self-debug.
\newblock In \emph{The Twelfth International Conference on Learning Representations}, 2024{\natexlab{a}}.

\bibitem[Chen et~al.(2024{\natexlab{b}})Chen, Hu, Zhi, Han, Deng, and Yin]{chen2024chatunitestframeworkllmbasedtest}
Yinghao Chen, Zehao Hu, Chen Zhi, Junxiao Han, Shuiguang Deng, and Jianwei Yin.
\newblock Chatunitest: A framework for llm-based test generation, 2024{\natexlab{b}}.

\bibitem[Chu et~al.(2023)Chu, Zhao, Weber, Li, and Wermter]{chu2023accelerating}
Kun Chu, Xufeng Zhao, Cornelius Weber, Mengdi Li, and Stefan Wermter.
\newblock Accelerating reinforcement learning of robotic manipulations via feedback from large language models.
\newblock \emph{arXiv preprint arXiv:2311.02379}, 2023.

\bibitem[DeepSeek-AI(2025)]{deepseekai2025deepseekr1incentivizingreasoningcapability}
DeepSeek-AI.
\newblock Deepseek-r1: Incentivizing reasoning capability in llms via reinforcement learning, 2025.

\bibitem[El-Kishky et~al.(2025)El-Kishky, Wei, Saraiva, Minaiev, Selsam, Dohan, Song, Lightman, Clavera, Pachocki, et~al.]{el2025competitive}
Ahmed El-Kishky, Alexander Wei, Andre Saraiva, Borys Minaiev, Daniel Selsam, David Dohan, Francis Song, Hunter Lightman, Ignasi Clavera, Jakub Pachocki, et~al.
\newblock Competitive programming with large reasoning models.
\newblock \emph{arXiv preprint arXiv:2502.06807}, 2025.

\bibitem[Forg{\'a}cs \& Kov{\'a}cs(2024)Forg{\'a}cs and Kov{\'a}cs]{forgacs2024modern}
Istv{\'a}n Forg{\'a}cs and Attila Kov{\'a}cs.
\newblock \emph{Modern software testing techniques}.
\newblock Springer, 2024.

\bibitem[Guo et~al.(2024)Guo, Okamura, and Dohi]{guo2024optimal}
Xiujing Guo, Hiroyuki Okamura, and Tadashi Dohi.
\newblock Optimal test case generation for boundary value analysis.
\newblock \emph{Software Quality Journal}, 32\penalty0 (2):\penalty0 543--566, 2024.

\bibitem[Guzu et~al.(2025)Guzu, Nicolae, Cucu, and Burileanu]{guzu2025large}
Alexandru Guzu, Georgian Nicolae, Horia Cucu, and Corneliu Burileanu.
\newblock Large language models for c test case generation: A comparative analysis.
\newblock \emph{Electronics}, 14\penalty0 (11):\penalty0 2284, 2025.

\bibitem[Han et~al.(2024)Han, Kim, Yoo, Lee, and won Hwang]{han2024archcodeincorporatingsoftwarerequirements}
Hojae Han, Jaejin Kim, Jaeseok Yoo, Youngwon Lee, and Seung won Hwang.
\newblock Archcode: Incorporating software requirements in code generation with large language models, 2024.

\bibitem[He et~al.(2025)He, Choi, Zhang, Ji, Zhou, Xu, Bercovich, Zhang, and Li]{he2025hardtestssynthesizinghighqualitytest}
Zhongmou He, Yee~Man Choi, Kexun Zhang, Jiabao Ji, Junting Zhou, Dejia Xu, Ivan Bercovich, Aidan Zhang, and Lei Li.
\newblock Hardtests: Synthesizing high-quality test cases for llm coding, 2025.

\bibitem[Hendrycks et~al.(2021)Hendrycks, Basart, Kadavath, Mazeika, Arora, Guo, Burns, Puranik, He, Song, et~al.]{hendrycks2021measuring}
Dan Hendrycks, Steven Basart, Saurav Kadavath, Mantas Mazeika, Akul Arora, Ethan Guo, Collin Burns, Samir Puranik, Horace He, Dawn Song, et~al.
\newblock Measuring coding challenge competence with apps.
\newblock \emph{arXiv preprint arXiv:2105.09938}, 2021.

\bibitem[Hou et~al.(2025)Hou, Zhang, Ji, Liu, Qian, Andreas, and Chang]{hou2025thinkprune0}
Bairu Hou, Yang Zhang, Jiabao Ji, Yujian Liu, Kaizhi Qian, Jacob Andreas, and Shiyu Chang.
\newblock Thinkprune: Pruning long chain-of-thought of llms via reinforcement learning.
\newblock \emph{arXiv preprint arXiv: 2504.01296}, 2025.

\bibitem[Hui et~al.(2024)Hui, Yang, Cui, Yang, Liu, Zhang, Liu, Zhang, Yu, Lu, Dang, Fan, Zhang, Yang, Men, Huang, Zheng, Miao, Quan, Feng, Ren, Ren, Zhou, and Lin]{hui2024qwen25codertechnicalreport}
Binyuan Hui, Jian Yang, Zeyu Cui, Jiaxi Yang, Dayiheng Liu, Lei Zhang, Tianyu Liu, Jiajun Zhang, Bowen Yu, Keming Lu, Kai Dang, Yang Fan, Yichang Zhang, An~Yang, Rui Men, Fei Huang, Bo~Zheng, Yibo Miao, Shanghaoran Quan, Yunlong Feng, Xingzhang Ren, Xuancheng Ren, Jingren Zhou, and Junyang Lin.
\newblock Qwen2.5-coder technical report, 2024.

\bibitem[Intellect(2025)]{primeintellect_synthetic1_2025}
Prime Intellect.
\newblock Synthetic-1: Scaling distributed synthetic data generation for verified reasoning.
\newblock \url{https://www.primeintellect.ai/blog/synthetic-1}, 2025.

\bibitem[Jain et~al.(2025)Jain, Han, Gu, Li, Yan, Zhang, Wang, Solar-Lezama, Sen, and Stoica]{jain2025livecodebench}
Naman Jain, King Han, Alex Gu, Wen-Ding Li, Fanjia Yan, Tianjun Zhang, Sida Wang, Armando Solar-Lezama, Koushik Sen, and Ion Stoica.
\newblock Livecodebench: Holistic and contamination free evaluation of large language models for code.
\newblock In \emph{The Thirteenth International Conference on Learning Representations}, 2025.

\bibitem[Ji et~al.(2025)Ji, Chen, Zhang, Kompella, Fan, Liu, and Chang]{ji2025collision0}
Jiabao Ji, Yongchao Chen, Yang Zhang, Ramana~Rao Kompella, Chuchu Fan, Gaowen Liu, and Shiyu Chang.
\newblock Collision- and reachability-aware multi-robot control with grounded llm planners.
\newblock \emph{arXiv preprint arXiv: 2505.20573}, 2025.

\bibitem[Jimenez et~al.(2024)Jimenez, Yang, Wettig, Yao, Pei, Press, and Narasimhan]{jimenez2024swebench}
Carlos~E Jimenez, John Yang, Alexander Wettig, Shunyu Yao, Kexin Pei, Ofir Press, and Karthik~R Narasimhan.
\newblock {SWE}-bench: Can language models resolve real-world github issues?
\newblock In \emph{The Twelfth International Conference on Learning Representations}, 2024.

\bibitem[Kimi(2025)]{kimiteam2025kimik15scalingreinforcement}
Team Kimi.
\newblock Kimi k1.5: Scaling reinforcement learning with llms, 2025.

\bibitem[Lambert et~al.(2025)Lambert, Morrison, Pyatkin, Huang, Ivison, Brahman, Miranda, Liu, Dziri, Lyu, Gu, Malik, Graf, Hwang, Yang, Bras, Tafjord, Wilhelm, Soldaini, Smith, Wang, Dasigi, and Hajishirzi]{lambert2025tulu3pushingfrontiers}
Nathan Lambert, Jacob Morrison, Valentina Pyatkin, Shengyi Huang, Hamish Ivison, Faeze Brahman, Lester James~V. Miranda, Alisa Liu, Nouha Dziri, Shane Lyu, Yuling Gu, Saumya Malik, Victoria Graf, Jena~D. Hwang, Jiangjiang Yang, Ronan~Le Bras, Oyvind Tafjord, Chris Wilhelm, Luca Soldaini, Noah~A. Smith, Yizhong Wang, Pradeep Dasigi, and Hannaneh Hajishirzi.
\newblock Tulu 3: Pushing frontiers in open language model post-training, 2025.

\bibitem[Le et~al.(2022)Le, Wang, Gotmare, Savarese, and Hoi]{le2022coderl0}
Hung Le, Yue Wang, Akhilesh~Deepak Gotmare, Silvio Savarese, and Steven C.~H. Hoi.
\newblock Coderl: Mastering code generation through pretrained models and deep reinforcement learning.
\newblock \emph{arXiv preprint arXiv: 2207.01780}, 2022.

\bibitem[Li et~al.(2025)Li, Tang, Wang, and Guo]{lipatchpilot}
Hongwei Li, Yuheng Tang, Shiqi Wang, and Wenbo Guo.
\newblock Patchpilot: A cost-efficient software engineering agent with early attempts on formal verification.
\newblock In \emph{Forty-second International Conference on Machine Learning}, 2025.

\bibitem[Li \& Yuan(2024)Li and Yuan]{li2024largelanguagemodelstest}
Kefan Li and Yuan Yuan.
\newblock Large language models as test case generators: Performance evaluation and enhancement, 2024.

\bibitem[Li et~al.(2023)Li, Fu, Zhang, Huang, Sun, Lyu, Liu, Jin, and Li]{li2023tacotopicsalgorithmiccode}
Rongao Li, Jie Fu, Bo-Wen Zhang, Tao Huang, Zhihong Sun, Chen Lyu, Guang Liu, Zhi Jin, and Ge~Li.
\newblock Taco: Topics in algorithmic code generation dataset, 2023.

\bibitem[Li et~al.(2022)Li, Choi, Chung, Kushman, Schrittwieser, Leblond, Eccles, Keeling, Gimeno, Dal~Lago, Hubert, Choy, de~Masson~d’Autume, Babuschkin, Chen, Huang, Welbl, Gowal, Cherepanov, Molloy, Mankowitz, Sutherland~Robson, Kohli, de~Freitas, Kavukcuoglu, and Vinyals]{Li_2022_alpha}
Yujia Li, David Choi, Junyoung Chung, Nate Kushman, Julian Schrittwieser, Rémi Leblond, Tom Eccles, James Keeling, Felix Gimeno, Agustin Dal~Lago, Thomas Hubert, Peter Choy, Cyprien de~Masson~d’Autume, Igor Babuschkin, Xinyun Chen, Po-Sen Huang, Johannes Welbl, Sven Gowal, Alexey Cherepanov, James Molloy, Daniel~J. Mankowitz, Esme Sutherland~Robson, Pushmeet Kohli, Nando de~Freitas, Koray Kavukcuoglu, and Oriol Vinyals.
\newblock Competition-level code generation with alphacode.
\newblock \emph{Science}, 378\penalty0 (6624):\penalty0 1092–1097, December 2022.
\newblock ISSN 1095-9203.
\newblock \doi{10.1126/science.abq1158}.

\bibitem[Lin et~al.(2025)Lin, Shen, Shang, Weston, and Nie]{lin2025learningsolveverifyselfplay}
Zi~Lin, Sheng Shen, Jingbo Shang, Jason Weston, and Yixin Nie.
\newblock Learning to solve and verify: A self-play framework for code and test generation, 2025.

\bibitem[Liu \& Zhang(2025)Liu and Zhang]{code-r1}
Jiawei Liu and Lingming Zhang.
\newblock Code-r1: Reproducing r1 for code with reliable rewards.
\newblock 2025.

\bibitem[Liu et~al.(2023)Liu, Xia, Wang, and ZHANG]{liu2023is}
Jiawei Liu, Chunqiu~Steven Xia, Yuyao Wang, and LINGMING ZHANG.
\newblock Is your code generated by chat{GPT} really correct? rigorous evaluation of large language models for code generation.
\newblock In \emph{Thirty-seventh Conference on Neural Information Processing Systems}, 2023.

\bibitem[Llama(2024)]{grattafiori2024llama3herdmodels}
Llama.
\newblock The llama 3 herd of models, 2024.

\bibitem[Luo et~al.(2025)Luo, Tan, Huang, Patel, Ariyak, Wu, Shi, Xin, Cai, Weber, Zhang, Li, Popa, and Stoica]{deepcoder2025}
Michael Luo, Sijun Tan, Roy Huang, Ameen Patel, Alpay Ariyak, Qingyang Wu, Xiaoxiang Shi, Rachel Xin, Colin Cai, Maurice Weber, Ce~Zhang, Li~Erran Li, Raluca~Ada Popa, and Ion Stoica.
\newblock Deepcoder: A fully open-source 14b coder at o3-mini level, 2025.
\newblock Notion Blog.

\bibitem[MatrixStudio(2025)]{matrixstudio_codeforces_python_2025}
MatrixStudio.
\newblock Codeforces python submissions.
\newblock \url{https://huggingface.co/datasets/MatrixStudio/Codeforces-Python-Submissions}, 2025.

\bibitem[OpenAI(2024)]{openai2024openaio1card}
OpenAI.
\newblock Openai o1 system card, 2024.

\bibitem[Penedo et~al.(2025)Penedo, Lozhkov, Kydlíček, Allal, Beeching, Lajarín, Gallouédec, Habib, Tunstall, and von Werra]{penedo2025codeforces}
Guilherme Penedo, Anton Lozhkov, Hynek Kydlíček, Loubna~Ben Allal, Edward Beeching, Agustín~Piqueres Lajarín, Quentin Gallouédec, Nathan Habib, Lewis Tunstall, and Leandro von Werra.
\newblock Codeforces.
\newblock \url{https://huggingface.co/datasets/open-r1/codeforces}, 2025.

\bibitem[Prasad et~al.(2025)Prasad, Stengel-Eskin, Chen, Khan, and Bansal]{prasad2025learninggenerateunittests}
Archiki Prasad, Elias Stengel-Eskin, Justin Chih-Yao Chen, Zaid Khan, and Mohit Bansal.
\newblock Learning to generate unit tests for automated debugging, 2025.

\bibitem[Puspitasari et~al.(2023)Puspitasari, Kurniasari, and Puspitasari]{puspitasari2023analysis}
TD~Puspitasari, AA~Kurniasari, and PSD Puspitasari.
\newblock Analysis and testing using boundary value analysis methods for geographic information system.
\newblock In \emph{IOP Conference Series: Earth and Environmental Science}, volume 1168, pp.\  012051. IOP Publishing, 2023.

\bibitem[Reid(1997)]{reid1997empirical}
Stuart~C Reid.
\newblock An empirical analysis of equivalence partitioning, boundary value analysis and random testing.
\newblock In \emph{Proceedings fourth international software metrics symposium}, pp.\  64--73. IEEE, 1997.

\bibitem[Shao et~al.(2024)Shao, Wang, Zhu, Xu, Song, Bi, Zhang, Zhang, Li, Wu, and Guo]{shao2024deepseekmathpushinglimitsmathematical}
Zhihong Shao, Peiyi Wang, Qihao Zhu, Runxin Xu, Junxiao Song, Xiao Bi, Haowei Zhang, Mingchuan Zhang, Y.~K. Li, Y.~Wu, and Daya Guo.
\newblock Deepseekmath: Pushing the limits of mathematical reasoning in open language models, 2024.

\bibitem[Sheng et~al.(2024)Sheng, Zhang, Ye, Wu, Zhang, Zhang, Peng, Lin, and Wu]{sheng2024hybridflow}
Guangming Sheng, Chi Zhang, Zilingfeng Ye, Xibin Wu, Wang Zhang, Ru~Zhang, Yanghua Peng, Haibin Lin, and Chuan Wu.
\newblock Hybridflow: A flexible and efficient rlhf framework.
\newblock 2024.

\bibitem[Sinha et~al.(2025)Sinha, Goel, Kumaraguru, Geiping, Bethge, and Prabhu]{sinha2025languagemodelsfalsifyevaluating}
Shiven Sinha, Shashwat Goel, Ponnurangam Kumaraguru, Jonas Geiping, Matthias Bethge, and Ameya Prabhu.
\newblock Can language models falsify? evaluating algorithmic reasoning with counterexample creation, 2025.

\bibitem[Touvron et~al.(2023)Touvron, Martin, Stone, Albert, Almahairi, Babaei, Bashlykov, Batra, Bhargava, Bhosale, Bikel, Blecher, Ferrer, Chen, Cucurull, Esiobu, Fernandes, Fu, Fu, Fuller, Gao, Goswami, Goyal, Hartshorn, Hosseini, Hou, Inan, Kardas, Kerkez, Khabsa, Kloumann, Korenev, Koura, Lachaux, Lavril, Lee, Liskovich, Lu, Mao, Martinet, Mihaylov, Mishra, Molybog, Nie, Poulton, Reizenstein, Rungta, Saladi, Schelten, Silva, Smith, Subramanian, Tan, Tang, Taylor, Williams, Kuan, Xu, Yan, Zarov, Zhang, Fan, Kambadur, Narang, Rodriguez, Stojnic, Edunov, and Scialom]{touvron2023llama2openfoundation}
Hugo Touvron, Louis Martin, Kevin Stone, Peter Albert, Amjad Almahairi, Yasmine Babaei, Nikolay Bashlykov, Soumya Batra, Prajjwal Bhargava, Shruti Bhosale, Dan Bikel, Lukas Blecher, Cristian~Canton Ferrer, Moya Chen, Guillem Cucurull, David Esiobu, Jude Fernandes, Jeremy Fu, Wenyin Fu, Brian Fuller, Cynthia Gao, Vedanuj Goswami, Naman Goyal, Anthony Hartshorn, Saghar Hosseini, Rui Hou, Hakan Inan, Marcin Kardas, Viktor Kerkez, Madian Khabsa, Isabel Kloumann, Artem Korenev, Punit~Singh Koura, Marie-Anne Lachaux, Thibaut Lavril, Jenya Lee, Diana Liskovich, Yinghai Lu, Yuning Mao, Xavier Martinet, Todor Mihaylov, Pushkar Mishra, Igor Molybog, Yixin Nie, Andrew Poulton, Jeremy Reizenstein, Rashi Rungta, Kalyan Saladi, Alan Schelten, Ruan Silva, Eric~Michael Smith, Ranjan Subramanian, Xiaoqing~Ellen Tan, Binh Tang, Ross Taylor, Adina Williams, Jian~Xiang Kuan, Puxin Xu, Zheng Yan, Iliyan Zarov, Yuchen Zhang, Angela Fan, Melanie Kambadur, Sharan Narang, Aurelien Rodriguez, Robert Stojnic, Sergey Edunov, and Thomas
  Scialom.
\newblock Llama 2: Open foundation and fine-tuned chat models, 2023.

\bibitem[Wang et~al.(2025{\natexlab{a}})Wang, Yang, Tian, Shen, and Wang]{wang2025co}
Yinjie Wang, Ling Yang, Ye~Tian, Ke~Shen, and Mengdi Wang.
\newblock Co-evolving llm coder and unit tester via reinforcement learning.
\newblock \emph{arXiv preprint arXiv:2506.03136}, 2025{\natexlab{a}}.

\bibitem[Wang et~al.(2025{\natexlab{b}})Wang, Liu, Sun, Li, and Shen]{wang2025codecontestshighqualitytestcase}
Zihan Wang, Siyao Liu, Yang Sun, Hongyan Li, and Kai Shen.
\newblock Codecontests+: High-quality test case generation for competitive programming, 2025{\natexlab{b}}.

\bibitem[Xia et~al.(2025)Xia, Shen, Wang, Liu, Sun, Wu, Hu, and Xu]{xia2025leetcodedatasettemporaldatasetrobust}
Yunhui Xia, Wei Shen, Yan Wang, Jason~Klein Liu, Huifeng Sun, Siyue Wu, Jian Hu, and Xiaolong Xu.
\newblock Leetcodedataset: A temporal dataset for robust evaluation and efficient training of code llms, 2025.

\bibitem[Xiong et~al.(2023)Xiong, Guo, and Chen]{xiong2023program}
Weimin Xiong, Yiwen Guo, and Hao Chen.
\newblock The program testing ability of large language models for code.
\newblock \emph{arXiv preprint arXiv:2310.05727}, 2023.

\bibitem[Yang et~al.(2025)Yang, Li, Yang, Zhang, Hui, Zheng, Yu, Gao, Huang, Lv, Zheng, Liu, Zhou, Huang, Hu, Ge, Wei, Lin, Tang, Yang, Tu, Zhang, Yang, Yang, Zhou, Zhou, Lin, Dang, Bao, Yang, Yu, Deng, Li, Xue, Li, Zhang, Wang, Zhu, Men, Gao, Liu, Luo, Li, Tang, Yin, Ren, Wang, Zhang, Ren, Fan, Su, Zhang, Zhang, Wan, Liu, Wang, Cui, Zhang, Zhou, and Qiu]{yang2025qwen3technicalreport}
An~Yang, Anfeng Li, Baosong Yang, Beichen Zhang, Binyuan Hui, Bo~Zheng, Bowen Yu, Chang Gao, Chengen Huang, Chenxu Lv, Chujie Zheng, Dayiheng Liu, Fan Zhou, Fei Huang, Feng Hu, Hao Ge, Haoran Wei, Huan Lin, Jialong Tang, Jian Yang, Jianhong Tu, Jianwei Zhang, Jianxin Yang, Jiaxi Yang, Jing Zhou, Jingren Zhou, Junyang Lin, Kai Dang, Keqin Bao, Kexin Yang, Le~Yu, Lianghao Deng, Mei Li, Mingfeng Xue, Mingze Li, Pei Zhang, Peng Wang, Qin Zhu, Rui Men, Ruize Gao, Shixuan Liu, Shuang Luo, Tianhao Li, Tianyi Tang, Wenbiao Yin, Xingzhang Ren, Xinyu Wang, Xinyu Zhang, Xuancheng Ren, Yang Fan, Yang Su, Yichang Zhang, Yinger Zhang, Yu~Wan, Yuqiong Liu, Zekun Wang, Zeyu Cui, Zhenru Zhang, Zhipeng Zhou, and Zihan Qiu.
\newblock Qwen3 technical report, 2025.

\bibitem[Yu et~al.(2025)Yu, Zhang, Zhu, Yuan, Zuo, Yue, Dai, Fan, Liu, Liu, Liu, Lin, Lin, Ma, Sheng, Tong, Zhang, Zhang, Zhang, Zhu, Zhu, Chen, Chen, Wang, Yu, Song, Wei, Zhou, Liu, Ma, Zhang, Yan, Qiao, Wu, and Wang]{yu2025dapoopensourcellmreinforcement}
Qiying Yu, Zheng Zhang, Ruofei Zhu, Yufeng Yuan, Xiaochen Zuo, Yu~Yue, Weinan Dai, Tiantian Fan, Gaohong Liu, Lingjun Liu, Xin Liu, Haibin Lin, Zhiqi Lin, Bole Ma, Guangming Sheng, Yuxuan Tong, Chi Zhang, Mofan Zhang, Wang Zhang, Hang Zhu, Jinhua Zhu, Jiaze Chen, Jiangjie Chen, Chengyi Wang, Hongli Yu, Yuxuan Song, Xiangpeng Wei, Hao Zhou, Jingjing Liu, Wei-Ying Ma, Ya-Qin Zhang, Lin Yan, Mu~Qiao, Yonghui Wu, and Mingxuan Wang.
\newblock Dapo: An open-source llm reinforcement learning system at scale, 2025.

\bibitem[Yuan et~al.(2024)Yuan, Lou, Liu, Ding, Wang, Chen, and Peng]{yuan2024manualtestsevaluatingimproving}
Zhiqiang Yuan, Yiling Lou, Mingwei Liu, Shiji Ding, Kaixin Wang, Yixuan Chen, and Xin Peng.
\newblock No more manual tests? evaluating and improving chatgpt for unit test generation, 2024.

\bibitem[Zeng et~al.(2025)Zeng, Jiang, Wang, Nie, Chen, and Chen]{zeng2025acecoderacingcoderrl}
Huaye Zeng, Dongfu Jiang, Haozhe Wang, Ping Nie, Xiaotong Chen, and Wenhu Chen.
\newblock Acecoder: Acing coder rl via automated test-case synthesis, 2025.

\bibitem[Zhang et~al.(2023{\natexlab{a}})Zhang, Li, Li, Li, and Jin]{zhang-etal-2023-self}
Kechi Zhang, Zhuo Li, Jia Li, Ge~Li, and Zhi Jin.
\newblock Self-edit: Fault-aware code editor for code generation.
\newblock In Anna Rogers, Jordan Boyd-Graber, and Naoaki Okazaki (eds.), \emph{Proceedings of the 61st Annual Meeting of the Association for Computational Linguistics (Volume 1: Long Papers)}, pp.\  769--787, Toronto, Canada, July 2023{\natexlab{a}}. Association for Computational Linguistics.

\bibitem[Zhang et~al.(2023{\natexlab{b}})Zhang, Wang, Xia, Wang, and Li]{zhang2023algosynthesizingalgorithmicprograms}
Kexun Zhang, Danqing Wang, Jingtao Xia, William~Yang Wang, and Lei Li.
\newblock Algo: Synthesizing algorithmic programs with llm-generated oracle verifiers, 2023{\natexlab{b}}.

\end{thebibliography}
